\documentclass[a4paper,11pt]{article}

\usepackage{epsfig}
\usepackage{hyperref}
\usepackage{a4}
\usepackage{latexsym, amssymb} 
\usepackage{amsmath}
\usepackage{comment}
\usepackage{cite}

\newcommand{\lapprox}{%
\mathrel{%
\setbox0=\hbox{$<$}
\raise0.6ex\copy0\kern-\wd0
\lower0.65ex\hbox{$\sim$}
}}
\newcommand{\gapprox}{%
\mathrel{%
\setbox0=\hbox{$>$}
\raise0.6ex\copy0\kern-\wd0
\lower0.65ex\hbox{$\sim$}
}}

\catcode`@=11 
\def \gsim{\mathrel{\mathpalette\@versim>}}
\def \lsim{\mathrel{\mathpalette\@versim<}}
\def \@versim#1#2{\lower0.4ex\vbox{\baselineskip\z@skip\lineskip\z@skip
     \lineskiplimit\z@\ialign{$\m@th#1\hfil##\hfil$%
     \crcr#2\crcr\sim\crcr}}}
\catcode`@=12 

\makeatletter
 
 \@addtoreset{equation}{section}
\makeatother

\begin{document}

\begin{center}

{\large\bf Jet azimuthal angle correlations in the production of a Higgs boson pair plus two jets at hadron colliders}\\[15mm]
Junya Nakamura\footnote{E-mail: junya.nakamura@uni-tuebingen.de}, Julien Baglio\footnote{E-mail: julien.baglio@uni-tuebingen.de}, \\ \bigskip
{\em Institute for Theoretical Physics,
 University of T\"ubingen, \\
 Auf der Morgenstelle 14,
 72076 T\"ubingen, Germany.}
\\[20mm] 
\end{center}

\begin{abstract} 

Azimuthal angle correlations of two jets in the process $pp\to HHjj$ are studied.
The loop induced $\mathcal{O}(\alpha_s^4 \alpha_{}^2)$ gluon fusion (GF) sub-process and the $\mathcal{O}(\alpha_{}^4)$ weak boson fusion (WBF) sub-process are considered. 
The GF sub-process exhibits strong correlations in the azimuthal angles $\phi_{1,2}^{}$ of the two jets measured from the production plane of the Higgs boson pair and the difference between these two angles $\phi_1^{}-\phi_2^{}$, and a very small correlation in the sum of them $\phi_1^{}+\phi_2^{}$. 
The impact of using a finite top mass $m_t^{}$ value on the correlations is found crucial. The transverse momentum of the Higgs boson can be used to enhance or suppress the correlations. The impact of a non-standard value for the triple Higgs self-coupling on the correlations is found much smaller than that on other observables, such as the invariant mass of the two Higgs bosons.
The peak shifts of the azimuthal angle distributions reflect the magnitude of parity violation in the $gg\to HH$ amplitude and the dependence of the distributions on parity violating phases is analytically clarified. The WBF sub-process also produces correlated distributions and it is found that they are not induced by the quantum effect of the intermediate weak bosons but mainly by a kinematic effect. 
This kinematic effect is a characteristic feature of the WBF sub-process and is not observed in the GF sub-process. 
It is found that the correlations are different in the GF and in the WBF sub-processes. As part of the process dependent information, they will be helpful in the analyses of both the GF sub-process and the WBF sub-process at the LHC.

\end{abstract}

\vskip 1 true cm

\newpage
\setcounter{footnote}{0}

\def\baselinestretch{1.5}
\tableofcontents


\section{Introduction}\label{sec:intro}

The discovery of a Higgs boson with a mass around 125 GeV in 2012 is the main discovery of Run I of the Large Hadron Collider (LHC)~\cite{Aad:2012tfa,Chatrchyan:2012xdj}. The study of its properties has started and until now they are compatible with the Standard Model (SM) hypothesis~\cite{Aad:2014eva,Khachatryan:2014kca,ATLAS-CONF-2015-008}.
To probe the mechanism of electroweak symmetry breaking (EWSB)~\cite{Higgs:1964ia,Englert:1964et,Higgs:1964pj,Guralnik:1964eu} directly one would want to measure the triple Higgs self-coupling that is one of the key parameters of the scalar potential. This is one of the main goals of the future high-luminosity LHC and the Future Circular Collider (FCC) in hadron--hadron mode, a potential 100 TeV proton--proton collider following the LHC at CERN (for reviews of the FCC physics potential, see refs.~\cite{Arkani-Hamed:2015vfh,Baglio:2015wcg}). In this view the production of a pair of Higgs bosons needs to be observed and has been extensively studied over the last years~\cite{Djouadi:1999gv,Djouadi:1999rca,Baur:2002rb,Baur:2002qd,Baur:2003gpa,Baur:2003gp,Dolan:2012rv,Baglio:2012np,Papaefstathiou:2012qe,Barr:2013tda,Barger:2013jfa,Dolan:2013rja,deFlorian:2013uza, deFlorian:2013jea, Goertz:2014qta,deLima:2014dta,Frederix:2014hta,Maltoni:2014eza, Azatov:2015oxa,Dolan:2015zja,deFlorian:2015moa} (see also Refs.~\cite{Barr:2014sga,Papaefstathiou:2015iba,Kotwal:2015rba} for studies at the FCC). 
The gluon fusion (GF) sub-process~\cite{Eboli:1987dy,Glover:1987nx, Dicus:1987ic,Plehn:1996wb} and the weak boson fusion (WBF) sub-process~\cite{Keung:1987nw, Eboli:1987dy, Dicus:1987ez, Dobrovolskaya:1990kx} are the two main sub-processes, see e.g. Ref.~\cite{Baglio:2014aka} for a review of SM studies. Both of the two sub-processes are sensitive to the triple Higgs self-coupling. In the WBF sub-process, we have access to the coupling between the two Higgs bosons and the two weak bosons, too. In Refs.~\cite{Dolan:2013rja,Dolan:2015zja} studies using the production of a pair of Higgs bosons plus two hadronic jets has been conducted, using both the GF and the WBF sub-processes. The main advantage of the latter sub-process is the fact that the theoretical uncertainties are under control~\cite{Baglio:2012np,Frederix:2014hta}, but the phenomenological studies suffer from the difficulty to separate the GF contributions from that of the WBF. \\

Azimuthal angle correlations of two jets produced together with heavy particles have been actively studied as a provider of important information about the heavy particles~\cite{Plehn:2001nj, Hankele:2006ma, Klamke:2007cu, Hagiwara:2009wt, Buckley:2010jv, Campanario:2010mi, Hagiwara:2013jp}. The correlations are induced by only a certain type of sub-processes, called vector boson fusion (VBF) sub-processes, in which a heavy object is produced by a fusion of two vector bosons emitted from incoming two coloured particles. The correlations arise from the quantum effect of the two fusing intermediate vector bosons~\cite{Dobrovolskaya:1990kx, Hagiwara:2009wt}~\footnote{The azimuthal angle correlations of two outgoing electrons in $e^+_{}e^-_{}$ collisions, induced by the quantum effect of two intermediate virtual photons, have been discussed a long time ago, see e.g.~\cite{Budnev:1974de}.}.
A set of cuts on the rapidity $y_{1,2}^{}$ of the two jets, $y_2^{} < 0 < y_1^{}$ and $y_1^{}-y_2^{}\gtrsim 3$ and an upper cut on the transverse momentum $p_T^{}$ of the two jets are therefore crucial for the two jets to show strong correlations if any, since they enhance contributions from VBF sub-processes~\cite{Hagiwara:2009wt}. These rapidity cuts are often called VBF cuts. \\

The azimuthal angle correlations of the two jets both in the GF sub-process and in the WBF sub-process of the single Higgs boson plus two jets production process $pp\to Hjj$~\cite{Dobrovolskaya:1990kx, Bambade:1993vw, Plehn:2001nj, Hankele:2006ma, Klamke:2007cu, Hagiwara:2009wt} are nowadays a common knowledge, have been studied in detail~\cite{DelDuca:2003ba,Figy:2004pt,Campbell:2006xx,DelDuca:2006hk,Andersen:2007mp,Ciccolini:2007ec,Andersen:2008gc,Bredenstein:2008tm,Nason:2009ai,Andersen:2010zx,Campanario:2013mga,Campanario:2014oua} and applied in many phenomenological studies.
However, the azimuthal angle correlations of the two jets in the Higgs boson pair plus two jets production process $pp\to HHjj$ have not been studied thoroughly. 
To our knowledge, the correlations in the GF sub-process, which is an one-loop induced $\mathcal{O}(\alpha_s^4 \alpha_{}^2)$ process at leading order (LO), have not been studied in the literature. One of the reasons may be that event generations are still challenging even with an advanced calculation technique. For the GF sub-process in the process $pp\to Hjj$, the approach of using the effective interactions between the Higgs boson and gluons is known to work quite well as long as the $p_T^{}$ of the jets are small enough~\cite{DelDuca:2001eu, DelDuca:2001fn}. The azimuthal angle correlation after the VBF cuts can also be described correctly~\cite{DelDuca:2001fn}. Therefore event generations can be easily performed with a tree-level event generator which implements the effective interactions. In contrast, for the GF sub-process in the process $pp\to HHjj$, the effective interaction approach is known not to work well in describing the distributions of several observables~\cite{Dolan:2013rja, Dolan:2015zja}. It is naively expected that this observation is the same for the azimuthal angle correlations. The process $pp\to HHjj$ at LO with the exact one loop amplitude has been calculated for the first time in ref.~\cite{Dolan:2013rja} and subsequently phenomenology is studied in ref.~\cite{Dolan:2015zja}. 
The fully automated event generation for one-loop induced processes is now available in {\tt MadGraph5\verb|_|aMC@NLO}~\cite{Alwall:2014hca, Hirschi:2015iia}. This achievement will activate further phenomenological studies of the process $pp\to HHjj$ including studies which use the azimuthal angle correlations. 
The azimuthal angle correlations of the two jets in the WBF sub-process, which is an $\mathcal{O}(\alpha_{}^4)$ process at LO, have been studied in~\cite{Dobrovolskaya:1990kx}. There, only the azimuthal angle difference of the two jets is studied as an azimuthal angle observable. \\

\begin{figure}[t]
\centering
\includegraphics[scale=0.62]{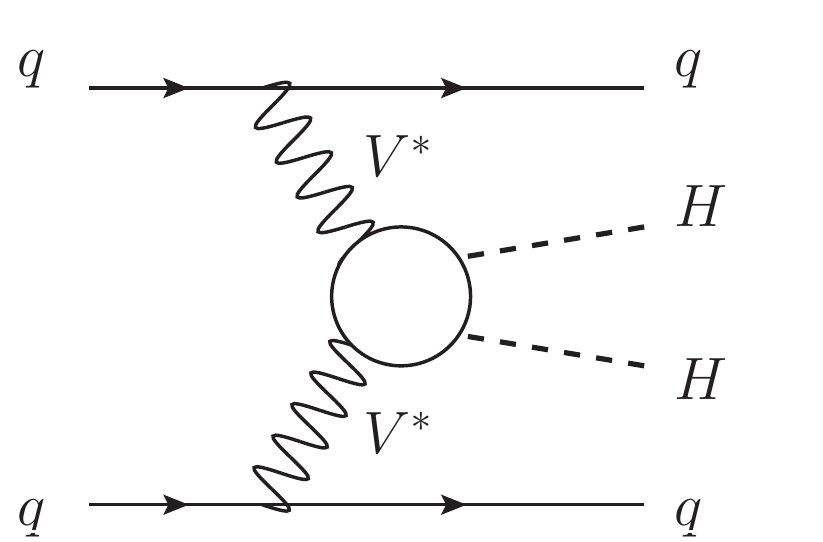}
\includegraphics[scale=0.62]{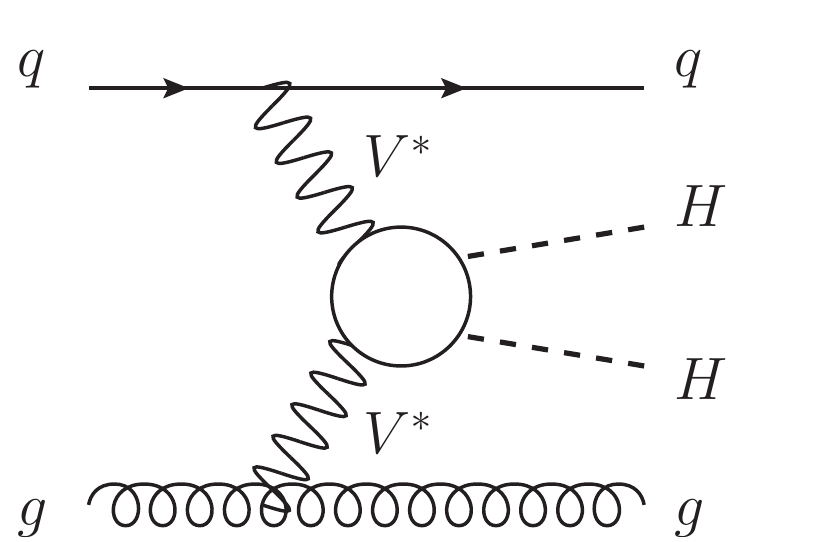}
\includegraphics[scale=0.62]{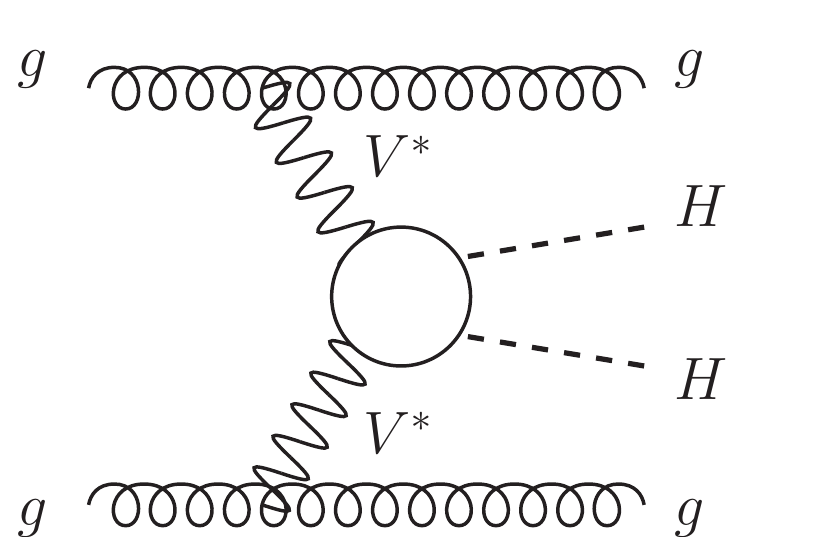}
\caption{\small Feynman diagrams calculated in this paper. The circles denote all the Feynman diagrams contributing to the process $V^*_{}V^*_{} \to HH$ at LO.}
\label{figure:diagram-123}
\end{figure}

In this paper, we study the azimuthal angle correlations of the two jets in the process $pp \to HHjj$. 
Instead of considering all of the sub-processes contributing to the process $pp \to HHjj$, we consider only the VBF sub-processes:
\begin{subequations}\label{eq:sub-processes}
\begin{align}
& qq \to qq V^*_{}V^*_{} \to qq HH\ \ \ \ \     ( V = W, Z, g ), \label{eq:sub-processes-1}\\
& qg \to qg V^*_{}V^*_{} \to qg HH\ \ \ \ \     ( V = g ), \label{eq:sub-processes-2}\\
& gg \to gg V^*_{}V^*_{} \to gg HH\ \ \ \ \     ( V = g ), \label{eq:sub-processes-3}
\end{align}
\end{subequations}
where $q$ denotes a light quark or a light antiquark, $g$ denotes a gluon and $V^*_{}$ denotes an intermediate off-shell vector boson. The two Higgs bosons are produced by a fusion of two vector bosons emitted from incoming two coloured particles. 
More precisely, we calculate the amplitudes contributed from only $t$-channel Feynman diagrams shown in Figure~\ref{figure:diagram-123}~\footnote{All pictures in this paper are drawn by using the program {\tt jaxodraw}~\cite{Binosi:2003yf}.}. We call them VBF amplitudes and VBF diagrams, respectively, in this paper. 
The circles denote all the Feynman diagrams contributing to the process $V^*_{}V^*_{} \to HH$ at LO. When the two virtual vector bosons are weak bosons (WBF sub-process), there are four tree-level Feynman diagrams, and when the two virtual vector bosons are gluons (GF sub-process), there are eight one-loop Feynman diagrams. We calculate these LO diagrams explicitly. \\

Our calculation is not exact but an approximated one.
In ref.~\cite{Hagiwara:2009wt} it has been demonstrated that the inclusive cross section and the azimuthal angle correlations contributed only from the VBF diagrams are in good agreement with those contributed from all the diagrams, when the two jets are required to satisfy the VBF cuts and an upper $p_T^{}$ cut, in the processes $pp\to \Phi jj$, where $\Phi= H, A, G$ denotes the Higgs boson, a parity-odd Higgs boson and a spin-2 massive graviton, respectively. 
This observation indicates that a result based only on VBF diagrams is a good approximation to the exact LO result as long as the two jets satisfy the VBF cuts and an upper $p_T^{}$ cut appropriately. 
Therefore our approach of calculating only the VBF amplitude is the simple and appropriate approach, when we study the azimuthal angle correlations (Let us remind that the correlations are induced only by the VBF sub-processes.).
The same approach has been applied to the processes $pp\to Q\bar{Q}jj$, where $Q$ denotes the top quark or the bottom quark, in ref.~\cite{Hagiwara:2013jp} and the validity of this approach is again observed.  
In order to check the validity of this approach in the GF sub-process by ourselves, we have numerically compared our cross section formula in which the loop-running top quark mass is set quite large (large $m_t^{}$ limit) and the exact LO result generated by {\tt MadGraph5\verb|_|aMC@NLO}~\cite{Alwall:2014hca} which implements the effective interactions between gluons and the Higgs bosons (infinite $m_t^{}$ limit).
We have found a good agreement between the two results both in the inclusive cross section and in the distributions of several observables including azimuthal angle observables.
From these observations, we are confident that our calculation of the azimuthal angle correlations is a good approximation to the exact LO result.\\

In order to measure the azimuthal angle correlations, we study four observables: $\phi_1^{}$, $\phi_2^{}$, $\Delta \phi = \phi_1^{} - \phi_2^{}$ and $\phi_+^{} = \phi_1^{} + \phi_2^{}$, where $\phi_{1,2}^{}$ are azimuthal angles of the two jets measured from the production plane of the Higgs boson pair ($\Delta \phi$ is irrelevant to the production plane as is clear from its definition).
$\Delta \phi$ is the observable which is sensitive to the property of the Higgs boson in the process $pp\to Hjj$~\cite{Plehn:2001nj, Hankele:2006ma, Klamke:2007cu, Hagiwara:2009wt, Campanario:2010mi} and thus has been the subject of many studies. The processes $pp\to Gjj$~\cite{Hagiwara:2009wt} and $pp\to Q\bar{Q}jj$~\cite{Hagiwara:2013jp} exhibit strong correlations in $\phi_+^{}$. To our knowledge, correlations in $\phi_{1,2}^{}$ have not been addressed in any hadronic process in the literature. 
Our explicit findings can be summarised as follows. The GF sub-process has strong correlations in $\Delta \phi$ and $\phi_{1,2}^{}$, and the $p_T^{}$ of the Higgs boson can be an useful measure to enhance or suppress these correlations. Using the finite $m_t^{}$ value is important to produce the correlations correctly. 
Violation of the parity invariance of the $gg\to HH$ amplitude appears as the peak shifts of the correlations.  
The impact of a non-standard value for the triple Higgs self-coupling on the correlations is smaller than that on other observables, such as the invariant mass of the two Higgs bosons, of the inclusive process $pp\to HH$.
The correlation in $\phi_+^{}$ is negligibly small in most every case. 
The WBF sub-process produces correlated distributions in all of the azimuthal angle observables and they are not induced by the quantum effect of the intermediate weak bosons but mainly by a kinematic effect. This kinematic effect is a characteristic feature of the WBF sub-process and is not observed in the GF sub-process. The impact of a non-standard value for the triple Higgs self-coupling on the correlations is not significant in the WBF sub-process, too. The correlations in the GF and WBF sub-processes are found to be different. \\

The paper is organised as follows. In Section~\ref{sec:Analytic_formula}, we perform a calculation of the VBF amplitudes. Since it can be shown that the azimuthal angle correlations arise from the interference of amplitudes with various helicities of the two intermediate vector bosons~\cite{Dobrovolskaya:1990kx, Hagiwara:2009wt}, we employ a helicity amplitude technique. Our calculation is performed based on the method presented in ref.~\cite{Hagiwara:2009wt}. A full analytic set of helicity amplitudes is presented. Since the material in Section~\ref{sec:Analytic_formula} is rather technical, the reader who is interested in only the results may skip this section. 
In Section~\ref{sec:correlation}, a detailed study of the azimuthal angle correlations is presented. First, we discuss the GF sub-process in Section~\ref{sec:gf-process}. The squared VBF amplitude for the four sub-processes eq.~(\ref{eq:sub-processes}) is given in a compact form. 
This analytic formula is found to be quite useful in making expectations of the correlations. 
The correlations in different kinematic regions of the two Higgs bosons and those in non-standard values for the Higgs triple self-coupling are studied. 
The impact of parity violation in the $gg\to HH$ amplitude on the correlations is also studied.
Next, we discuss the WBF sub-process in Section~\ref{sec:wbf-process}. The squared VBF amplitude is given in a simple form by keeping only the dominant terms. 
The correlations in non-standard values for the Higgs triple self-coupling are studied. In Section~\ref{sec:conclusion} we summarise our findings and give some comments.

\section{Helicity amplitudes for the process {\boldmath $pp \to V^*_{}V^*_{}jj \to HHjj$}}\label{sec:Analytic_formula}

In this section we present a full analytic set of helicity amplitudes contributed from the vector boson fusion (VBF) diagrams. Our calculation is based on the method presented in ref.~\cite{Hagiwara:2009wt}. We present a more complete discussion on the treatment of the intermediate off-shell gluons in the VBF diagrams. 
In addition, we discuss the importance of an appropriate choice of gauge-fixing vectors for the polarisation vectors of the external gluons. 
We believe that the above two remarks provide sufficient justification for repeating some of the calculations of ref.~\cite{Hagiwara:2009wt}.

\subsection{VBF amplitudes}\label{sec:derivation}

\begin{figure}[t]
\centering
\includegraphics[scale=0.6]{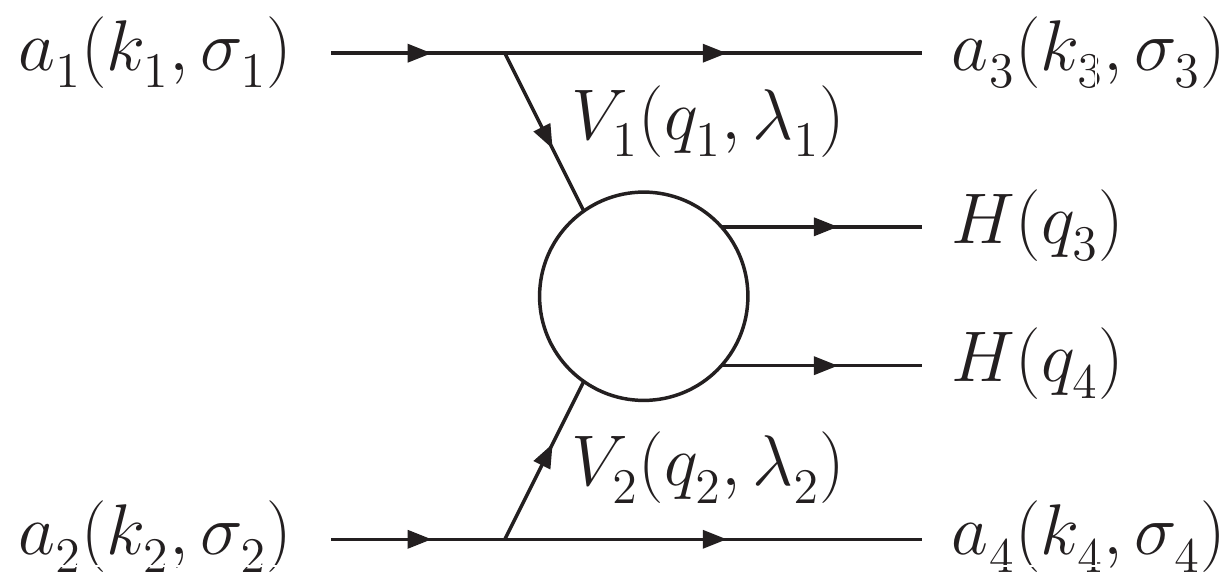}
\caption{\small The assignment of the four-momenta and helicities of each particle for the VBF sub-processes.}
\label{figure:vbf-kinematics}
\end{figure}

We first introduce a common set of kinematic variables
\begin{align}
a_1^{}\bigl(k_1^{}, \sigma_1^{}\bigr) + a_2^{}\bigl(k_2^{}, \sigma_2^{}\bigr) & \to a_3^{}\bigl(k_3^{}, \sigma_3^{}\bigr) + a_4^{}\bigl(k_4^{}, \sigma_4^{}\bigr) + V_1^{}\bigl(q_1^{}, \lambda_1^{}\bigr) + V_2^{}\bigl(q_2^{}, \lambda_2^{}\bigr) \nonumber \\
& \to a_3^{}\bigl(k_3^{}, \sigma_3^{}\bigr) + a_4^{}\bigl(k_4^{}, \sigma_4^{}\bigr) + H\bigl(q_3^{}\bigr) + H\bigl(q_4^{}\bigr),\label{eq:mom-assign}
\end{align}
where $a_{1,2,3,4}^{}$ can be quarks, antiquarks or gluons, $V_{1,2}^{}$ are intermediate off-shell vector bosons, $H$ denotes the Higgs boson and the four-momentum $k_i^{}$ and helicity $\sigma_i^{}$ of each particle are shown. This assignment for the VBF sub-processes is more apparent in Figure~\ref{figure:vbf-kinematics}. 
The external particles take helicities $\sigma_i^{}=\pm 1$~\footnote{We define the helicity operator for a two-component spinor by $\vec{p} \cdot \vec{\sigma} / |\vec{p}|$ with the Pauli matrices $\vec{\sigma}$, so a quark also takes $\sigma = \pm 1$ in our notation. Sometimes we simply write $\sigma = \pm$ instead of $\sigma = \pm 1$, and do the same for $\lambda$.} and the intermediate vector bosons take helicities $\lambda_i^{}=\pm 1, 0$. Colour indices are suppressed. The VBF helicity amplitude can be expressed as follows:
\begin{align}
{\cal M}_{\sigma_1^{} \sigma_2^{}}^{\sigma_3^{} \sigma_4^{}} = J_{V_1^{}a_1^{}a_3^{}}^{\mu_1^{\prime}}\bigl( k_1^{}, k_3^{}; \sigma_1^{}, \sigma_3^{}\bigl) J_{V_2^{}a_2^{}a_4^{}}^{\mu_2^{\prime}}\bigl( k_2^{}, k_4^{}; \sigma_2^{}, \sigma_4^{}\bigl) D_{\mu_1^{\prime}\mu_1^{}}^{V_1^{}}\bigl( q_1^{} \bigr) D_{\mu_2^{\prime}\mu_2^{}}^{V_2^{}}\bigl( q_2^{} \bigr) \Gamma^{\mu_1^{}\mu_2^{}}_{HHV_1^{}V_2^{}}\bigl(q_1^{},q_2^{},q_3^{},q_4^{}\bigr),\label{eq:full-amplitudes-1}
\end{align}
where $J_{V_i^{}a_i^{}a_{i+2}^{}}^{\mu_i^{\prime}}( k_i^{}, k_{i+2}^{}; \sigma_i^{}, \sigma_{i+2}^{})$ is a current involving the off-shell vector boson $V_i^{}$ and the two external quarks, antiquarks or gluons, $D_{\mu_i^{\prime}\mu_i^{}}^{V_i^{}}( q_i^{})$ is the $V_i^{}$ propagator and $\Gamma^{\mu_1^{}\mu_2^{}}_{HHV_1^{}V_2^{}}\bigl(q_1^{},q_2^{},q_3^{},q_4^{}\bigr)$ is the tensor amplitude for the process $V_1^{}V_2^{} \to HH$. When we denote a helicity amplitude in this paper, we show only helicity indices. \\

For the weak boson fusion (WBF) sub-process, we have $(V_1^{}, V_2^{})=(W^+_{}, W^-_{})$, $(W^-_{}, W^+_{})$ and $(Z, Z)$, and only the VBF diagram shown in Figure~\ref{figure:diagram-123} (left) contributes to the VBF amplitude. The VBF amplitude is gauge invariant on its own. 
This is apparent from the fact that only the VBF diagram exists in some of the WBF sub-processes, for instance there are no other diagrams describing the sub-process $u d\to u d H H$.
We choose the unitary gauge for the weak boson propagator,
\begin{subequations}\label{eq:prop-weak}
\begin{align}
D_{\mu_{}^{\prime}\mu}^{V}\bigl( q \bigr)
& = \biggl( - g_{\mu_{}^{\prime}\mu}^{} + \frac{ q_{\mu_{}^{\prime}}^{} q_{\mu}^{} }{m_{V}^2} \biggr) D_{V}^{}\bigl( q \bigr), \label{eq:uni-gauge-proj}\\
D_{V}^{}\bigl( q \bigr) & = \bigl( q_{}^2 - m_{V}^2 \bigr)^{-1}_{}.\label{eq:prop-fact}
\end{align}
\end{subequations}
We express the projector part of the above propagator in terms of polarisation vectors in the helicity basis $(\lambda=\pm 1, 0, s)$:
\begin{align}
- g^{\mu_{}^{\prime}\mu}_{} + \frac{ q_{}^{\mu_{}^{\prime}} q_{}^{\mu} }{m_{V}^2} & = - g^{\mu_{}^{\prime}\mu}_{} + \frac{ q_{}^{\mu_{}^{\prime}} q_{}^{\mu} }{q_{}^2} - \biggl( 1 - \frac{q_{}^2}{m_{V}^2} \biggr) \frac{ q_{}^{\mu_{}^{\prime}} }{ \sqrt{-q_{}^2} } \frac{ q_{}^{\mu} }{ \sqrt{-q_{}^2} } \nonumber \\
& = \sum_{\lambda=\pm 1, 0} \bigl(-1\bigr)^{\lambda+1}_{} \epsilon^{\mu_{}^{\prime}}_{}\bigl(q, \lambda \bigr)^*_{} \epsilon^{\mu}_{}\bigl(q, \lambda\bigr) 
- \biggl( 1 - \frac{q_{}^2}{m_{V}^2} \biggr) 
\epsilon^{\mu_{}^{\prime}}_{}\bigl(q, \lambda=s \bigr)^*_{} \epsilon^{\mu}_{}\bigl(q, \lambda=s \bigr). \label{eq:projector-1}
\end{align}
The $\lambda=s$ component $\epsilon^{\mu}_{}(q, \lambda=s) = q_{}^{\mu} / \sqrt{-q_{}^2}$ is the scalar part of a virtual weak boson and vanishes when it couples with a light quark current. As a result, the following replacement is possible when the external quarks are assumed to be massless: 
\begin{align}
- g^{\mu_{}^{\prime}\mu}_{} + \frac{ q_{}^{\mu_{}^{\prime}} q_{}^{\mu} }{m_{V}^2} \to \sum_{\lambda=\pm 1, 0} \bigl(-1\bigr)^{\lambda+1}_{} \epsilon^{\mu_{}^{\prime}}_{}\bigl(q, \lambda \bigr)^*_{} \epsilon^{\mu}_{}\bigl(q, \lambda\bigr). \label{eq:projector-2}
\end{align}
The polarisation vectors $\epsilon^{\mu}_{}(q_i^{}, \lambda_i^{}=\pm,0)$ will be defined later once the kinematics of the weak bosons is fixed. After that, one can confirm eq.~(\ref{eq:projector-1}) explicitly. By inserting the identity eq.~(\ref{eq:projector-2}) into the VBF helicity amplitude in eq. (\ref{eq:full-amplitudes-1}), it can be expressed as a product of three helicity amplitudes:
\begin{align}
{\cal M}_{\sigma_1^{} \sigma_2^{}}^{\sigma_3^{} \sigma_4^{}} = D_{V_1^{}}^{}\bigl( q_1^{} \bigr) D_{V_2^{}}^{}\bigl( q_2^{} \bigr) 
\sum_{\lambda_1^{}=\pm, 0} \sum_{\lambda_2^{}=\pm, 0} 
\bigl( {\cal J}_{a_1^{}a_3^{}}^{V_1^{}} \bigr)_{\sigma_1^{} \sigma_3^{}}^{\lambda_1^{}}
\bigl( {\cal J}_{a_2^{}a_4^{}}^{V_2^{}} \bigr)_{\sigma_2^{} \sigma_4^{}}^{\lambda_2^{}}
\bigl( {\cal M}_{V_1^{}V_2^{}}^{HH} \bigr)_{\lambda_1^{} \lambda_2^{}}^{} \label{eq:full-amplitudes-2}
\end{align}
with
\begin{subequations}\label{eq:full-amplitudes-2-set}
\begin{align}
\bigl( {\cal J}_{a_i^{}a_{i+2}^{}}^{V_i^{}} \bigr)_{\sigma_i^{} \sigma_{i+2}^{}}^{\lambda_i^{}} & = \bigl(-1\bigr)^{\lambda_i^{}+1}_{} J_{V_i^{}a_i^{}a_{i+2}^{}}^{\mu_i^{\prime}}\bigl( k_i^{}, k_{i+2}^{}; \sigma_i^{}, \sigma_{i+2}^{}\bigl) \epsilon_{\mu_i^{\prime}}^{}\bigl(q_i^{}, \lambda_i^{}\bigr)^*_{},\label{eq:current-amp-1} \\
\bigl( {\cal M}_{V_1^{}V_2^{}}^{HH} \bigr)_{\lambda_1^{} \lambda_2^{}}^{}
& = \epsilon_{\mu_1^{}}^{}\bigl(q_1^{}, \lambda_1^{}\bigr) 
  \epsilon_{\mu_2^{}}^{}\bigl(q_2^{}, \lambda_2^{}\bigr) 
\Gamma^{\mu_1^{}\mu_2^{}}_{HHV_1^{}V_2^{}}\bigl(q_1^{},q_2^{},q_3^{},q_4^{}\bigr).\label{eq:core-amp-1}
\end{align} 
\end{subequations}
Eq.~(\ref{eq:current-amp-1}) represents a helicity amplitude for the splitting process $a_i^{} \to a_{i+2}^{} V_i^{}$, where $V_i^{}$ is off-shell. This will be derived in Section~\ref{sec:hel-split}. 
Eq.~(\ref{eq:core-amp-1}) represents a helicity amplitude for the process $V_1^{} V_2^{} \to HH$, where $V_1^{}$ and $V_2^{}$ are off-shell. This will be presented in Section~\ref{sec:hel-coreprocess}. \\

The gluon fusion (GF) sub-process $(V_1^{}, V_2^{})=(g, g)$ is more complicated than the WBF sub-process. The VBF amplitude for the $qq$ initiated sub-process $(a_1^{}, a_2^{})=(q, q)$ is gauge invariant on its own, the reason being the same as for the WBF amplitude.
If we choose the Feynman-'t Hooft gauge for a gluon propagator, the projector part of the propagator is:
\begin{align}
- g^{\mu_{}^{\prime}\mu}_{} & = - g^{\mu_{}^{\prime}\mu}_{} + \frac{ q_{}^{\mu_{}^{\prime}} q_{}^{\mu} }{q_{}^2} - \frac{ q_{}^{\mu_{}^{\prime}} }{ \sqrt{-q_{}^2} } \frac{ q_{}^{\mu} }{ \sqrt{-q_{}^2} } \nonumber \\
& = \sum_{\lambda=\pm1, 0} \bigl(-1\bigr)^{\lambda+1}_{} \epsilon^{\mu_{}^{\prime}}_{}\bigl(q, \lambda \bigr)^*_{} \epsilon^{\mu}_{}\bigl(q, \lambda\bigr) 
- \epsilon^{\mu_{}^{\prime}}_{}\bigl(q, \lambda=s \bigr)^*_{} \epsilon^{\mu}_{}\bigl(q, \lambda=s \bigr). \label{eq:gluon-projector-1}
\end{align}
The $\lambda=s$ component $\epsilon^{\mu}_{}(q, \lambda=s) = q_{}^{\mu} / \sqrt{-q_{}^2}$ again vanishes when it couples with a quark current. As a result, the following replacement is possible without any approximation:
\begin{align}
- g^{\mu_{}^{\prime}\mu}_{} \to \sum_{\lambda=\pm 1, 0} \bigl(-1\bigr)^{\lambda+1}_{} \epsilon^{\mu_{}^{\prime}}_{}\bigl(q, \lambda \bigr)^*_{} \epsilon^{\mu}_{}\bigl(q, \lambda \bigr). \label{eq:gluon-projector-2}
\end{align}
Therefore, for the $qq$ initiated sub-process, we can arrive at the same expression as in eq.~(\ref{eq:full-amplitudes-2}). In ref.~\cite{Hagiwara:2009wt}, the above procedure and thus the expression eq.~(\ref{eq:full-amplitudes-2}) is used not only for the $qq$ initiated sub-process but also for the $qg$ and $gg$ initiated sub-processes $(a_1^{}, a_2^{})=(q, g), (g, g)$~\footnote{The $\lambda=s$ component $\epsilon^{\mu}_{}(q, \lambda=s) = q_{}^{\mu} / \sqrt{-q_{}^2}$ vanishes when it couples to a gluon current, too.}. We point out that this approach does not necessarily calculate the off-shell effects of the intermediate gluons correctly for the $qg$ and $gg$ initiated sub-processes and only introduces unnecessary complications in the amplitude calculation. 
Since the off-shell gluon amplitude $( {\cal M}_{gg}^{HH} )_{\lambda_1^{} \lambda_2^{}}^{}$ is not enhanced in the on-shell limit of the gluons, we can always expand the off-shell gluon amplitude around the on-shell limit. 
To make our discussion simpler, let us consider an amplitude which involves only one off-shell gluon. The off-shell gluon amplitude can be expanded as
\begin{align}
{\cal M}_g^{}\bigl(Q \bigr) & = {\cal M}_g^{}\bigl(Q=0 \bigr) + c_1^{} Q + c_2^{} Q^2_{} + \cdots, \nonumber \\
c_n^{} & = \frac{1}{n!}\frac{d^n_{} {\cal M}_g^{}\bigl(Q\bigr) }{(dQ)^n_{}}\biggr|_{Q \to 0}^{}, \label{eq:expand-1}
\end{align}
where $Q$ is the virtuality of the gluon and the first term in the right hand side (RHS) of the first equation is the on-shell amplitude, which is gauge invariant.
As we will see in Section~\ref{sec:hel-split}, the amplitude for the splitting process in eq.~(\ref{eq:current-amp-1}), where an off-shell gluon with virtuality $Q$ is emitted, has an overall factor of $Q$. 
By considering this factor and $Q^{-2}$ in the propagator factor of the off-shell gluon, we find the following term in a VBF amplitude:
\begin{align}
 Q \times \frac{1}{Q^2_{}} \times {\cal M}_g^{}\bigl(Q \bigr) = \frac{{\cal M}_g^{}\bigl(Q=0 \bigr)}{Q} + c_1^{} + c_2^{} Q + \cdots.\label{eq:expand-2}
\end{align}
While the first term in the RHS of the above equation is gauge invariant, the rest terms are generally dependent on a gauge-fixing choice for the off-shell gluon. 
As we have mentioned above, for the $qq$ initiated sub-process, the VBF amplitude is gauge invariant on its own and hence not only the first term in the RHS of eq.~(\ref{eq:expand-2}) but also the other terms as a whole are gauge invariant. 
This is not the case for the $qg$ and $gg$ initiated sub-processes. 
For the $qg$ and $gg$ initiated sub-processes, the second and higher terms in the RHS of eq.~(\ref{eq:expand-2}), which are not enhanced in the on-shell limit, become gauge invariant only after contributions from other diagrams are also included. 
Therefore, in our method where only the VBF diagrams are calculated, we can calculate the off-shell effect of the intermediate gluons correctly only for the $qq$ initiated sub-process. 
For the $qg$ and $gg$ initiated sub-processes, therefore, we take the following approach: Completely ignore the off-shell effect of the intermediate gluons and look at only a kinematic region where the virtualities of the gluons are small ($Q\to 0$). This is possible because the off-shell effect in the amplitude will not be essential as long as we look at the small virtuality region, as is clear from eq.~(\ref{eq:expand-2}). 
The unitarity condition gives the following equation for a gluon propagator in its on-shell limit $q_{}^2 \to 0$:
\begin{subequations}\label{eq:gluon-propagator}
\begin{align}
D_{\mu_{}^{\prime}\mu}^{g}\bigl( q \bigr)
& = D_{g}^{}\bigl( q \bigr) 
\sum_{\lambda=\pm} \epsilon_{\mu_{}^{\prime}}^{}\bigl(q, \lambda \bigr)^*_{} \epsilon_{\mu}^{}\bigl(q, \lambda \bigr), \label{eq:gluon-proj}\\
D_{g}^{}\bigl( q \bigr) & = \bigl( q_{}^2 \bigr)^{-1}_{}.\label{eq:gluon-prop-fact}
\end{align}
\end{subequations}
By using this, we can express the VBF amplitude for the GF sub-process as a product of three helicity amplitudes in the same way that we do for the WBF sub-process and the $qq$ initiated GF sub-process in eq.~(\ref{eq:full-amplitudes-2}). 
The differences from the $qq$ initiated GF sub-process are (1) the $\lambda=0$ component $\epsilon^{\mu}_{}(q, \lambda=0)$ in eq.~(\ref{eq:full-amplitudes-2}) is neglected, (2) the off-shell gluon amplitude $( {\cal M}_{V_1^{}V_2^{}}^{HH} )_{\lambda_1^{} \lambda_2^{}}^{}$ is replaced by the on-shell gluon amplitude in which $q_1^2 = 0$ and $q_2^2 = 0$. Note that these two approximations are nothing less than the two fundamental ingredients of the equivalent photon approximation (EPA)~\cite{Williams:1934ad, vonWeizsacker:1934nji}. As a rule of using the EPA, we should clarify the kinematic region where the approximated VBF amplitude is a good approximation to the exact VBF amplitude. For the process $pp \to HHjj$, it should be $-q_1^2 < \hat{s}$ and $-q_2^2 < \hat{s}$, where $\hat{s} =M_{HH}^2$. 
We use the EPA not only for the $qg$ and $gg$ initiated sub-processes but also for the $qq$ initiated sub-process, so that we can consistently use the on-shell gluon amplitude $( {\cal M}_{V_1^{}V_2^{}}^{HH} )_{\lambda_1^{} \lambda_2^{}}^{}$.
It should be noted that, while we use the on-shell gluon amplitude for the process $gg\to HH$, the amplitude for an off-shell gluon emission in eq.~(\ref{eq:current-amp-1}) is still calculated without ignoring the off-shell effect of the gluon ($q_{1,2}^2 \ne 0$), as in other applications of the EPA.

\subsection{Helicity amplitudes for the splitting processes}\label{sec:hel-split}

We derive the helicity amplitude (\ref{eq:current-amp-1}) for the splitting processes. We use the chiral representation for Dirac matrices. Since we neglect the mass of the external quarks, the helicity of the quark is equal to its chirality and the helicity of the antiquark is opposite to its chirality. As a result, the helicity of $a_i^{}$ is always equal to that of $a_{i+2}^{}$ ($\sigma_i^{}=\sigma_{i+2}^{}$) for the quark and antiquark splitting processes. Otherwise, the amplitude vanishes. 
By introducing one common current $\hat{J}_i^{\mu}$, we can express the quark and antiquark currents $J_{V_i^{}a_i^{}a_{i+2}^{}}^{\mu}\bigl( k_i^{}, k_{i+2}^{}; \sigma_i^{}, \sigma_{i+2}^{}\bigl)$ in a compact manner:
\begin{subequations}
\begin{align}
J_{V_i^{}a_i^{}a_{i+2}^{}}^{\mu} \bigl( k_i^{}, k_{i+2}^{}; \sigma_i^{}, \sigma_{i+2}^{} \bigr) & = g_{\sigma_i^{}}^{V_i^{}a_i^{}a_{i+2}^{}} 
\hat{J}_i^{\mu} \bigl( k_i^{}, k_{i+2}^{}; \sigma_i^{}, \sigma_{i+2}^{} \bigr), \label{eq:quark-current} \\
\hat{J}_i^{\mu} \bigl( k_i^{}, k_{i+2}^{}; \sigma_i^{}, \sigma_{i+2}^{} \bigr) & =
u^{\dagger}_{} \bigl( k_{i+2}^{}, \sigma_{i+2}^{} \bigr)_{\alpha} \sigma^{\mu}_{\alpha} u\bigl(k_{i}^{}, \sigma_i^{}\bigr)_{\alpha}^{}
\ \delta_{\sigma_i^{} \sigma_{i+2}^{}}^{} \ \delta_{\sigma_i^{} \alpha}^{}
, \label{eq:quark-current-common} 
\end{align}
\end{subequations}
where $u(k,\sigma)_{\alpha}^{}$ represents the two-component Weyl $u$-spinor with its four-momentum $k$, helicity $\sigma$ and chirality $\alpha$ ($=\pm 1$), and $\sigma^{\mu}_{\pm} = (1, \pm \vec{\sigma})$ with the Pauli matrices $\vec{\sigma}$. Note that we can use the above $\hat{J}_i^{\mu}$ for the antiquark current, too, since
\begin{align}
v^{\dagger}_{}\bigl(k_{i}^{}, \sigma_i^{}\bigr)_{\alpha} \sigma^{\mu}_{\alpha} v\bigl(k_{i+2}^{}, \sigma_{i+2}^{}\bigr)_{\alpha}^{} = u^{\dagger}_{}\bigl(k_{i+2}^{}, \sigma_{i+2}^{}\bigr)_{-\alpha} \sigma^{\mu}_{-\alpha} u\bigl(k_{i}^{}, \sigma_i^{}\bigr)_{-\alpha}^{}.
\end{align}
The couplings between quarks and vector bosons relevant to our study are summarised as follows:
\begin{align}
&g_{\pm}^{gqq}  = g_{\pm}^{g\bar{q}\bar{q}} = g_s^{} t^{a}_{}, \nonumber \\
&g_{-}^{W^+_{}ud}  = g_{+}^{W^+_{}\bar{d}\bar{u}} = \frac{g_w^{}}{\sqrt{2}} \bigl( V_{ud}^{} \bigr)^*_{}, \nonumber \\
&g_{-}^{W^-_{}du}  = g_{+}^{W^-_{}\bar{u}\bar{d}} = \frac{g_w^{}}{\sqrt{2}} V_{ud}^{}, \nonumber \\
&g_{+}^{Zqq}  = g_{-}^{Z\bar{q}\bar{q}} = g_z^{} \bigl( - Q_q^{} \sin^2{\theta_w} \bigr), \nonumber \\
&g_{+}^{Z\bar{q}\bar{q}}  = g_{-}^{Zqq} = g_z^{} \bigl( T_{q}^3 - Q_q^{} \sin^2{\theta_w} \bigr),
\end{align}
where $g_s^{}$ is the QCD coupling, $t^a_{}$ is the generator of the SU(3) group, $g_w^{} = e / \sin{\theta_w^{}}$, $g_z^{} = e /(\sin{\theta_w^{}}\cos{\theta_w^{}})$ with $\theta_w^{}$ being the electroweak mixing angle and $e$ being the proton charge, $V_{ud}^{}$ is an element of the CKM matrix, $Q_q^{}$ is the electric charge of the quark in unit of $e$, $T^3_u=1/2$ and $T^3_d=-1/2$.
The gluon current involving three gluons is also expressed by eq.~(\ref{eq:quark-current}) with
\begin{subequations}
\begin{align}
 g^{ggg}_{\pm} & = g_s^{} f^{abc}_{}, \\
\hat{J}_i^{\mu} \bigl( k_i^{}, k_{i+2}^{}; \sigma_i^{}, \sigma_{i+2}^{} \bigr) & = \bigl( k_i^{} + q_i^{} \bigr) \cdot \epsilon ( k_{i+2}^{}, \sigma_{i+2}^{} )^*_{} \ \epsilon^{\mu}_{}\bigl( k_i^{}, \sigma_i^{} \bigr) 
+
\bigl( - q_i^{} + k_{i+2}^{} \bigr) \cdot \epsilon \bigl( k_{i}^{}, \sigma_{i}^{} \bigr) \ \epsilon^{\mu}\bigl( k_{i+2}^{}, \sigma_{i+2}^{} \bigr)^*_{} \nonumber \\
&+
\epsilon \bigl( k_{i}^{}, \sigma_{i}^{} \bigr) \cdot \epsilon \bigl( k_{i+2}^{}, \sigma_{i+2}^{} \bigr)^*_{} \ \bigl( - k_{i+2}^{} - k_i^{} \bigr)^{\mu}_{}, \label{eq:gluon-current}
\end{align}
\end{subequations}
where $f^{abc}_{}$ is the structure constant of the SU(3) group. \\

\begin{figure}[t]
\centering
\includegraphics[scale=0.55]{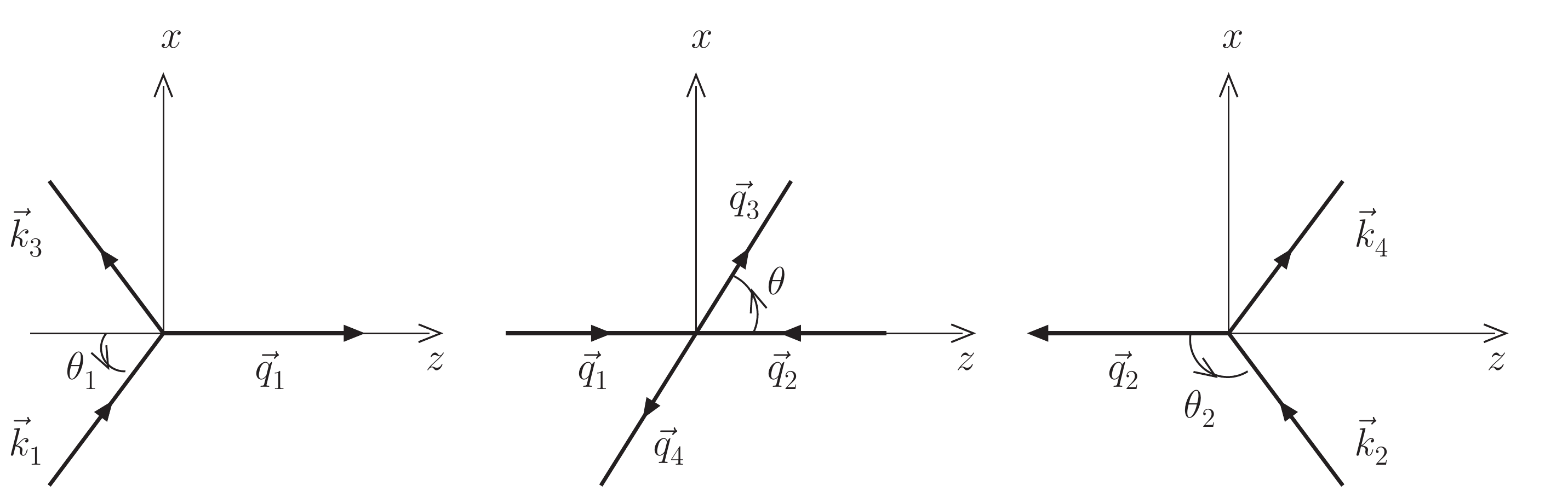}
\caption{\small The coordinate systems in the $q_1^{}$ Breit frame (left), the c.m. frame of the two colliding vector bosons (middle) and the $q_2^{}$ Breit frame (right). The $y$-axis points to us. The directions of the $z$-axis in the three coordinate systems are chosen common, so that the coordinate system in one of the frames can coincide with that in the other of the frames by a single boost along the $z$-axis.}
\label{figure:coordinate-system}
\end{figure}

Generally speaking, helicities are frame dependent and so are helicity amplitudes. When we calculate the VBF helicity amplitude ${\cal M}_{\sigma_1^{} \sigma_2^{}}^{\sigma_3^{} \sigma_4^{}}$ in eq.~(\ref{eq:full-amplitudes-2}), 
we must choose one frame at first and then define the four-momenta and the helicities of all the external particles in this particular frame. 
For our calculation we choose the centre-of-mass (c.m.) frame of the two intermediate vector bosons moving along the $z$-axis, which is shown in the middle of Figure~\ref{figure:coordinate-system}. We call it the VBF frame. Note that the production plane of the two Higgs bosons coincides with the plane of the $x$-$z$ axes. 
All of the three helicity amplitudes in the VBF helicity amplitude must be evaluated in the VBF frame. 
However, by using a property of helicities, we can justify a calculation of the helicity amplitude $( {\cal J}_{a_i^{}a_{i+2}^{}}^{V_i^{}} )_{\sigma_i^{} \sigma_{i+2}^{}}^{\lambda_i^{}}$ for the splitting processes in a different frame. 
Because the helicity of a massless quark is frame independent and furthermore the helicity of a massive vector boson is invariant under Lorentz boosts along its momentum direction as long as the boosts do not change the sign of its three-momentum~\footnote{This is also the case for an intermediate off-shell gluon.}, the helicity amplitude $( {\cal J}_{a_i^{}a_{i+2}^{}}^{V_i^{}})_{\sigma_i^{} \sigma_{i+2}^{}}^{\lambda_i^{}}$ for the quark splitting processes is invariant under Lorentz boosts along the $z$-axis from the VBF frame, as long as the boosts do not change the sign of the three-momentum of $V_i^{}$. 
By this property, it is justified to calculate the helicity amplitude $( {\cal J}_{a_1^{}a_{3}^{}}^{V_1^{}})_{\sigma_1^{} \sigma_{3}^{}}^{\lambda_1^{}}$ for the quark splitting process in the $q_1^{}$ Breit frame, to which $k_{1,3}^{}$ and $q_1^{}$ in the VBF frame can move by a single Lorentz boost along the negative direction of the $z$-axis. 
Similarly, it is justified to calculate the helicity amplitude $( {\cal J}_{a_2^{}a_{4}^{}}^{V_2^{}})_{\sigma_2^{} \sigma_{4}^{}}^{\lambda_2^{}}$ for the quark splitting process in the $q_2^{}$ Breit frame, to which $k_{2,4}^{}$ and $q_2^{}$ in the VBF frame can move by a single Lorentz boost along the positive direction of the $z$-axis. The $q_1^{}$ and $q_2^{}$ Breit frames are illustrated in the left and right of Figure~\ref{figure:coordinate-system}, respectively. 
A calculation of the helicity amplitude $( {\cal J}_{a_i^{}a_{i+2}^{}}^{V_i^{}})_{\sigma_i^{} \sigma_{i+2}^{}}^{\lambda_i^{}}$ for the gluon splitting process in the $q_i^{}$ Breit frame is also justified by appropriately choosing gauge-fixing vectors for the polarisation vectors of the external gluons. Although the helicity of a gluon is frame independent, the polarisation vectors of the gluon are dependent on their gauge-fixing vectors and hence the helicity amplitude for the gluon splitting process is in general frame dependent. 
However, if we choose the gauge fixing vectors in a way that their directions are invariant under Lorentz boosts along the $z$-axis, the helicity amplitude for the gluon splitting process also becomes invariant under the same boosts, again as long as the boosts do not change the sign of the three-momentum of the intermediate off-shell gluon. This can be simply achieved by choosing all the gauge-fixing vectors along the $z$-axis. \\

We parametrise the four-momenta $k_1^{}$, $k_3^{}$ and $q_1^{}$ in the $q_1^{}$ Breit frame as
\begin{align}\label{eq:breit1}
q_1^{\mu} & = k_1^{\mu} - k_3^{\mu} = \bigl( 0, \ 0,\ 0,\ Q_1^{} \bigr), \nonumber \\
k_1^{\mu} & = \frac{Q_1^{}}{2\cos{\theta_1^{}}} 
\bigl( 1,\  \sin{\theta_1^{}} \cos{\phi_1^{}},\ \sin{\theta_1^{}} \sin{\phi_1^{}}, \ \cos{\theta_1^{}} \bigr), \nonumber \\
k_3^{\mu} & = \frac{Q_1^{}}{2\cos{\theta_1^{}}} 
\bigl( 1,\  \sin{\theta_1^{}} \cos{\phi_1^{}},\  \sin{\theta_1^{}} \sin{\phi_1^{}},\ - \cos{\theta_1^{}} \bigr), 
\end{align}
where $Q_1^{} = \sqrt{- (q_1^{})^2_{}} > 0$, $0 < \theta_1^{} < \pi/2$ and $0 < \phi_1^{} < 2\pi$, and the four-momenta $k_2^{}$, $k_4^{}$ and $q_2^{}$ in the $q_2^{}$ Breit frame as
\begin{align}\label{eq:breit2}
q_2^{\mu} & = k_2^{\mu} - k_4^{\mu} = \bigl( 0, \ 0,\ 0,\ -Q_2^{} \bigr), \nonumber \\
k_2^{\mu} & = -\frac{Q_2^{}}{2\cos{\theta_2^{}}} 
\bigl( 1,\  \sin{\theta_2^{}} \cos{\phi_2^{}},\ \sin{\theta_2^{}} \sin{\phi_2^{}}, \ \cos{\theta_2^{}} \bigr), \nonumber \\
k_4^{\mu} & = -\frac{Q_2^{}}{2\cos{\theta_2^{}}} 
\bigl( 1,\  \sin{\theta_2^{}} \cos{\phi_2^{}},\  \sin{\theta_2^{}} \sin{\phi_2^{}},\ - \cos{\theta_2^{}} \bigr), 
\end{align}
where $Q_2^{} = \sqrt{- (q_2^{})^2_{}} > 0$, $\pi/2 < \theta_2^{} < \pi$ and $0 < \phi_2^{} < 2\pi$. We define polarisation vectors for the intermediate vector boson $V_1^{}$ in the $q_1^{}$ Breit frame by
\begin{align}
\epsilon^{\mu}_{} \bigl( q_1^{}, \lambda_1^{}= \pm \bigr)
&= \frac{1}{\sqrt{2}} \bigl( 0,\ - \lambda_1^{},\ -i,\ 0 \bigr),\nonumber \\
\epsilon^{\mu}_{} \bigl( q_1^{}, \lambda_1^{}= 0 \bigr)
&= \bigl( 1,\ 0,\ 0,\ 0 \bigr), \label{eq:pol-vect-1}
\end{align}
and those for the intermediate vector boson $V_2^{}$ in the $q_2^{}$ Breit frame by
\begin{align}
\epsilon^{\mu}_{} \bigl( q_2^{}, \lambda_2^{}= \pm \bigr)
&= \frac{1}{\sqrt{2}} \bigl( 0,\ - \lambda_2^{},\ i,\ 0 \bigr),\nonumber \\
\epsilon^{\mu}_{} \bigl( q_2^{}, \lambda_2^{}= 0 \bigr)
&= \bigl( 1,\ 0,\ 0,\ 0 \bigr). \label{eq:pol-vect-2}
\end{align}
It is easy to explicitly confirm that the above sets of the polarisation vectors satisfy the identity for the propagator of a weak boson in eq.~(\ref{eq:projector-1}). 
We use the same polarisation vectors $\epsilon^{\mu}_{} ( q_i^{}, \lambda_i^{}= \pm )$ for the intermediate weak bosons ($V_i^{}=W, Z$) and gluons ($V_i^{}=g$). 
As a result, the $\lambda_i^{}=\pm$ helicity amplitudes $( {\cal J}_{a_i^{}a_{i+2}^{}}^{V_i^{}})_{\sigma_i^{} \sigma_{i+2}^{}}^{\pm}$ for a weak boson emission from a quark and those for a gluon emission from a quark are the same, except for couplings. (Let us remind that the $\lambda=0$ components of the intermediate gluons are neglected.)\\

As we have discussed above, the gauge-fixing vectors for polarisation vectors of the external gluons should be chosen along the $z$-axis. For the polarisation vectors of the two external gluons in the $q_1^{}$ Breit frame, we choose a common vector $n_1^{\mu} = ( 1, 0, 0, -1)$, i.e. the light-cone axial gauge. With this choice, the polarisation vectors in the $q_1^{}$ Breit frame are
\begin{subequations}
\begin{align}
\epsilon^{\mu}_{}( k_1^{}, \sigma_1^{} ) = \frac{-\sigma_1^{}}{\sqrt{1+\cos{\theta_1^{}}}}
\biggl( 
\sin{\frac{\theta_1^{}}{2}} e^{i\sigma_1^{}\phi_1^{}}_{},
\ \cos{\frac{\theta_1^{}}{2}},
\ i \sigma_1^{} \cos{\frac{\theta_1^{}}{2}},
\ - \sin{\frac{\theta_1^{}}{2}} e^{i\sigma_1^{}\phi_1^{}}_{}
\biggr), \\
\epsilon^{\mu}_{}( k_3^{}, \sigma_3^{} ) = \frac{-\sigma_3^{}}{\sqrt{1-\cos{\theta_1^{}}}}
\biggl( 
\cos{\frac{\theta_1^{}}{2}} e^{i\sigma_3^{}\phi_1^{}}_{},
\ \sin{\frac{\theta_1^{}}{2}},
\ i \sigma_3^{} \sin{\frac{\theta_1^{}}{2}},
\ - \cos{\frac{\theta_1^{}}{2}} e^{i\sigma_3^{}\phi_1^{}}_{}
\biggr).
\end{align}
\end{subequations}
Similarly, a vector $n_2^{\mu} = ( 1, 0, 0, 1)$ is commonly chosen for the polarisation vectors of the two external gluons in the $q_2^{}$ Breit frame, then the polarisation vectors in the $q_2^{}$ Breit frame are
\begin{subequations}
\begin{align}
\epsilon^{\mu}_{}( k_2^{}, \sigma_2^{} ) = \frac{\sigma_2^{}}{\sqrt{1-\cos{\theta_2^{}}}}
\biggl( 
\cos{\frac{\theta_2^{}}{2}},
\ \sin{\frac{\theta_2^{}}{2}} e^{i\sigma_2^{}\phi_2^{}}_{},
\ -i \sigma_2^{} \sin{\frac{\theta_2^{}}{2}} e^{i\sigma_2^{}\phi_2^{}}_{},
\ \cos{\frac{\theta_2^{}}{2}}
\biggr), \\
\epsilon^{\mu}_{}( k_4^{}, \sigma_4^{} ) = \frac{\sigma_4^{}}{\sqrt{1+\cos{\theta_2^{}}}}
\biggl( 
\sin{\frac{\theta_2^{}}{2}},
\ \cos{\frac{\theta_2^{}}{2}} e^{i\sigma_4^{}\phi_2^{}}_{},
\ -i \sigma_4^{} \cos{\frac{\theta_2^{}}{2}} e^{i\sigma_4^{}\phi_2^{}}_{},
\ \sin{\frac{\theta_2^{}}{2}}
\biggr). 
\end{align}
\end{subequations}
It is easy to confirm that the above polarisation vectors with their gauge fixing vectors satisfy the unitarity condition for an on-shell gluon:
\begin{align}
-g^{\mu\nu}_{}+\frac{k^{\mu}_{} n^{\nu}_{} + n^{\mu}_{} k^{\nu}_{} }{k \cdot n}
=
\sum_{\sigma = \pm1} \epsilon^{\mu}_{}\bigl(k, \sigma \bigr)^*_{} \epsilon^{\nu}_{}\bigl(k, \sigma \bigr).
\end{align}
\\

With our preparations up to now, we can easily derive the helicity amplitude $( {\cal J}_{a_i^{}a_{i+2}^{}}^{V_i^{}})_{\sigma_i^{} \sigma_{i+2}^{}}^{\lambda_i^{}}$ for the quark splitting process and that for the gluon splitting process in the Breit frames. Following ref.~\cite{Hagiwara:2009wt}, we write the amplitudes as
\begin{align}
\bigl( {\cal J}_{a_i^{}a_{i+2}^{}}^{V_i^{}} \bigr)_{\sigma_i^{} \sigma_{i+2}^{}}^{\lambda_i^{}} = \sqrt{2}\ g_{\sigma_i^{}}^{V_i^{}a_i^{}a_{i+2}^{}}\ Q_i^{}\ {\cal \hat{J}}_{i\ \sigma_i^{} \sigma_{i+2}^{}}^{\ \lambda_i^{}}. \label{eq:common-amp-short}
\end{align}
The coupling factor is already defined in eq.~(\ref{eq:quark-current}). The common amplitude ${\cal \hat{J}}_{i\ \sigma_i^{} \sigma_{i+2}^{}}^{\ \lambda_i^{}}$ is summarised in Table~\ref{table:current-amplitude}. 
Note that we adopt the phase convention for the two-component Weyl spinors developed in refs.~\cite{Hagiwara:1985yu, Hagiwara:1988pp}. Since we use the same phase convention for the spinors and the same gauge fixing vectors for the external gluons with ref.~\cite{Hagiwara:2009wt}, the amplitudes in Table~\ref{table:current-amplitude} are consistent with those in tables 1 and 2 of ref.~\cite{Hagiwara:2009wt} including the common overall phase.

\begin{table}[t]
\centering
{\renewcommand\arraystretch{1.6}
\begin{tabular}{|c|c||c|c|}
\hline
\multicolumn{2}{|c||}{ ${\cal \hat{J}}_{1\ \sigma_1^{} \sigma_3^{}}^{\ \lambda_1^{}} ( q_1^{} \to q_3^{} V_1^* )$ } &
\multicolumn{2}{|c|}{ ${\cal \hat{J}}_{2\ \sigma_2^{} \sigma_4^{}}^{\ \lambda_2^{}} ( q_2^{} \to q_4^{} V_2^* )$ } \\ 
\hline 
${\cal \hat{J}}_{1\ \sigma \sigma }^{\ \sigma }$ & 
$ \mbox{\large $\frac{ \sigma }{2 \cos{\theta_1^{}} }$} \bigl( 1 + \cos{\theta_1^{}} \bigr)\ e^{-i\sigma \phi_1^{}}_{}$ &
${\cal \hat{J}}_{2\ \sigma \sigma }^{\ \sigma }$ & 
$ \mbox{\large $\frac{ - \sigma }{2 \cos{\theta_2^{}} }$} \bigl( 1 - \cos{\theta_2^{}} \bigr)\ e^{+i\sigma \phi_2^{}}_{}$ \\
${\cal \hat{J}}_{1\ \sigma \sigma }^{\ -\sigma }$ & 
$ \mbox{\large $\frac{ -\sigma }{2 \cos{\theta_1^{}} }$} \bigl( 1 - \cos{\theta_1^{}} \bigr)\ e^{+i\sigma \phi_1^{}}_{}$ &
${\cal \hat{J}}_{2\ \sigma \sigma }^{\ -\sigma }$ & 
$ \mbox{\large $\frac{  \sigma }{2 \cos{\theta_2^{}} }$} \bigl( 1 + \cos{\theta_2^{}} \bigr)\ e^{-i\sigma \phi_2^{}}_{}$ 
\\
${\cal \hat{J}}_{1\ \sigma \sigma }^{\ 0 }$ & 
$ \mbox{\large $\frac{ -1 }{2\cos{\theta_1^{}} }$} \sqrt{2} \sin{\theta_1^{}} $ &
${\cal \hat{J}}_{2\ \sigma \sigma }^{\ 0 }$ & 
$ \mbox{\large $\frac{ 1 }{2\cos{\theta_2^{}} }$} \sqrt{2} \sin{\theta_2^{}} $ \\
\hline
\hline
\multicolumn{2}{|c||}{ ${\cal \hat{J}}_{1\ \sigma_1^{} \sigma_3^{}}^{\ \lambda_1^{}} ( g_1^{} \to g_3^{} g_1^* )$ } &
\multicolumn{2}{|c|}{ ${\cal \hat{J}}_{2\ \sigma_2^{} \sigma_4^{}}^{\ \lambda_2^{}} ( g_2^{} \to g_4^{} g_2^* )$ } \\ 
\hline 
${\cal \hat{J}}_{1\ \sigma \sigma }^{\ \sigma }$ & 
$ \mbox{\large $\frac{ \sigma }{2 \sin{\theta_1^{}} \cos{\theta_1^{}} }$} \bigl( 1 + \cos{\theta_1^{}} \bigr)^2_{}\ e^{-i\sigma \phi_1^{}}_{}$ &
${\cal \hat{J}}_{2\ \sigma \sigma }^{\ \sigma }$ & 
$ \mbox{\large $\frac{ -\sigma }{2 \sin{\theta_2^{}} \cos{\theta_2^{}} }$} \bigl( 1 - \cos{\theta_2^{}} \bigr)^2_{}\ e^{+i\sigma \phi_2^{}}_{}$  \\
${\cal \hat{J}}_{1\ \sigma \sigma }^{\ -\sigma }$ & 
$ \mbox{\large $\frac{ -\sigma }{2 \sin{\theta_1^{}} \cos{\theta_1^{}} }$} \bigl( 1 - \cos{\theta_1^{}} \bigr)^2_{}\ e^{+i\sigma \phi_1^{}}_{}$ &
${\cal \hat{J}}_{2\ \sigma \sigma }^{\ -\sigma }$ & 
$ \mbox{\large $\frac{  \sigma }{2 \sin{\theta_2^{}} \cos{\theta_2^{}} }$} \bigl( 1 + \cos{\theta_2^{}} \bigr)^2_{}\ e^{-i\sigma \phi_2^{}}_{}$ \\
${\cal \hat{J}}_{1\ \sigma -\sigma }^{\ \sigma }$ & 
$ \mbox{\large $\frac{ -\sigma }{2 \sin{\theta_1^{}} \cos{\theta_1^{}} }$} 4 \cos^2_{}{\theta_1^{}} \ e^{+i\sigma \phi_1^{}}_{}$ &
${\cal \hat{J}}_{2\ \sigma -\sigma }^{\ \sigma }$ & 
$ \mbox{\large $\frac{  \sigma }{2 \sin{\theta_2^{}} \cos{\theta_2^{}} }$} 4 \cos^2_{}{\theta_2^{}} \ e^{+i\sigma \phi_2^{}}_{}$ \\
\hline
\end{tabular}
\caption{\small The helicity amplitudes ${\cal \hat{J}}_{i\ \sigma_i^{} \sigma_{i+2}^{}}^{\ \lambda_i^{}}$ defined in eq.~(\ref{eq:common-amp-short}) in the $q_1^{}$ Breit frame (left row) and the $q_2^{}$ Breit frame (right row). The amplitudes of an off-shell vector boson (weak boson or gluon) emission from a quark are given in the upper part and those of an off-shell gluon emission from a gluon are given in the lower part.}
\label{table:current-amplitude}
}
\end{table}

\subsection{Helicity amplitudes for the processes $VV \to HH$}\label{sec:hel-coreprocess}

Finally we present the helicity amplitudes for the processes $V_1^{} V_2^{} \to HH$, $( {\cal M}_{V_1^{}V_2^{}}^{HH} )_{\lambda_1^{} \lambda_2^{}}^{}$ defined in eq.~(\ref{eq:core-amp-1}). As we have mentioned in Section~\ref{sec:hel-split}, we evaluate the amplitudes in the VBF frame. We parametrise the four-momenta $q_1^{}$, $q_2^{}$, $q_3^{}$ and $q_4^{}$ in the VBF frame as
\begin{align}
q_1^{\mu} &= \biggl( \frac{1}{2\sqrt{\hat{s}}}\bigl(\hat{s} - Q_1^2 + Q_2^2\bigr),\ 0,\ 0,\ \sqrt{\frac{1}{4\hat{s}}\bigl(\hat{s} - Q_1^2 + Q_2^2\bigr)^2_{} + Q_1^2}\ \biggr), \nonumber \\
q_2^{\mu} &= \biggl( \frac{1}{2\sqrt{\hat{s}}}\bigl(\hat{s} + Q_1^2 - Q_2^2\bigr),\ 0,\ 0,\ -\sqrt{\frac{1}{4\hat{s}}\bigl(\hat{s} - Q_1^2 + Q_2^2\bigr)^2_{} + Q_1^2}\ \biggr),\nonumber \\
q_3^{\mu} &= \frac{\sqrt{\hat{s}}}{2} \bigl(1,\ \beta \sin{\theta},\ 0,\ \beta\cos{\theta} \bigr), \nonumber \\
q_4^{\mu} &= \frac{\sqrt{\hat{s}}}{2} \bigl(1,\ -\beta \sin{\theta},\ 0,\ -\beta\cos{\theta} \bigr), \label{eq:VBF-four-momenta}
\end{align}
where $\hat{s}$ is the c.m. energy squared $\hat{s}=(q_1^{}+q_2^{})^2_{}$ and $\beta = \sqrt{1-4m_H^2/\hat{s}}$ with $m_H^{}$ being the Higgs boson mass. The polarisation vectors for the vector bosons $V_{1,2}^{}$ in the VBF frame can be simply obtained by boosting those in the $q_{1,2}^{}$ Breit frames (eqs.~(\ref{eq:pol-vect-1}) and (\ref{eq:pol-vect-2})) to the VBF frame along the $z$-axis:
\begin{align}
\epsilon^{\mu}_{} \bigl( q_1^{}, \lambda_1^{}= \pm \bigr)
&= \frac{1}{\sqrt{2}} \bigl( 0,\ - \lambda_1^{},\ -i,\ 0 \bigr), \nonumber \\
\epsilon^{\mu}_{} \bigl( q_1^{}, \lambda_1^{}= 0 \bigr)
&= \frac{1}{Q_1^{}} \bigl(\ q_1^3,\ 0,\ 0,\ q_1^0\ \bigr), \label{eq:pol-vect-3}
\end{align}
and
\begin{align}
\epsilon^{\mu}_{} \bigl( q_2^{}, \lambda_2^{}= \pm \bigr)
&= \frac{1}{\sqrt{2}} \bigl( 0,\ - \lambda_2^{},\ i,\ 0 \bigr),\nonumber \\
\epsilon^{\mu}_{} \bigl( q_2^{}, \lambda_2^{}= 0 \bigr)
&= \frac{1}{Q_2^{}} \bigl(\ -q_2^3,\ 0,\ 0,\ -q_2^0\ \bigr). \label{eq:pol-vect-4}
\end{align}
Needless to say, the $\lambda_{1,2}^{} = \pm 1$ components remain the same after the boost. The helicity amplitude for the WBF process $V_1^{}V_2^{}\to HH$, where $(V_1^{}, V_2^{}) = (W^+_{}, W^-_{})$, $(W^-_{}, W^+_{})$ or $(Z, Z)$, is given by~\cite{Dobrovolskaya:1990kx}
\begin{align}
\bigl( {\cal M}_{V_1^{}V_2^{}}^{HH} \bigr)_{\lambda_1^{} \lambda_2^{}}^{}
 = 
\frac{4}{\sqrt{2}} G_F^{} m_{V_1^{}}^2 
\epsilon^{\mu_1^{}}_{}\bigl(q_1^{}, \lambda_1^{}\bigr) 
  \epsilon^{\mu_2^{}}_{}\bigl(q_2^{}, \lambda_2^{}\bigr) 
& \biggl\{ g_{\mu_1^{}\mu_2^{}}^{} 
+ 3 m_H^2 \lambda_h^{} D_H^{} \bigl( q_1^{} + q_2^{} \bigr) g_{\mu_1^{}\mu_2^{}}^{}  \nonumber \\
& - 2 m_{V_1^{}}^2 \Bigl[ D_{\mu_1^{}\mu_2^{}}^{V_1^{}}\bigl( q_1^{} - q_3^{} \bigr) + D_{\mu_1^{}\mu_2^{}}^{V_1^{}}\bigl( q_1^{} - q_4^{} \bigr) \Bigr] 
\biggr\}, 
\end{align}
where $G_F^{}$ is the Fermi constant and $\lambda_h^{}$ is the factor re-scaling the triple Higgs self-coupling: $\lambda_{HHH}^{}=\lambda_h^{} \lambda_{HHH}^{SM}$, where $\lambda_{HHH}^{SM}=3m_H^2/v$ with $v^{-2}_{}=\sqrt{2}G_F^{}$. The standard model predicts $\lambda_h^{}=1$. 
The notation for the propagators is defined in eq.~(\ref{eq:prop-weak}). By using the four-momenta and the polarisation vectors defined above, we can obtain the off-shell weak boson amplitude.\\

As we have discussed in Section~\ref{sec:derivation}, we use the on-shell gluon amplitude for the GF process $gg \to HH$. Using the notation given in the appendix of ref.~\cite{Plehn:1996wb}, we write the amplitude as
\begin{align}
\bigl( {\cal M}_{gg}^{HH} \bigr)_{\lambda_1^{} \lambda_2^{}}^{}
 = \frac{G_F^{}\alpha_s^{} \hat{s} }{2\sqrt{2}\pi} \delta_{b_1^{} b_2^{}}^{}
\epsilon^{\mu_1^{}}_{}\bigl(q_1^{}, \lambda_1^{}\bigr) 
  \epsilon^{\mu_2^{}}_{}\bigl(q_2^{}, \lambda_2^{}\bigr) 
\biggl\{
\Bigl[
3m_H^2 \lambda_h^{}  D_H^{} \bigl( q_1^{} + q_2^{} \bigr) F_{\triangle}^{} + F_{\square} \Bigr] A_{1 \mu_1^{} \mu_2^{}}^{} + 
G_{\square}^{} A_{2 \mu_1^{} \mu_2^{}}^{} 
\biggr\}, \label{eq:amp-gghh}
\end{align}
where $b_{1,2}^{}$ are the colour indices of the gluons, and $F_{\triangle}^{}$, $F_{\square}^{}$ and $G_{\square}^{}$ are form factors~\cite{Plehn:1996wb} consisting of the scalar loop functions after tensor reduction and $A_{i \mu_1 \mu_2}$ are the tensor structures of the process. Not only we neglect the $\lambda_{1,2}^{}=0$ components of the gluons, but we also set $Q_1^{}=Q_2^{}=0$ in the four-momenta in eq.~(\ref{eq:VBF-four-momenta}). 
Our definitions of the polarisation vectors $\epsilon^{\mu}_{} ( q_{i}^{}, \lambda_{i}^{}= \pm )$ in eqs.~(\ref{eq:pol-vect-3}) and (\ref{eq:pol-vect-4}) actually correspond to an axial gauge $n_{}^{\mu} = ( 1, 0, 0, -1)$ for an on-shell gluon $q_1^{\mu} \propto (1, 0, 0, 1)$ and to an axial gauge $n_{}^{\mu} = ( 1, 0, 0, 1)$ for an on-shell gluon $q_2^{\mu} \propto (1, 0, 0, -1)$, respectively.\\

In order to simplify further analyses, we introduce the following amplitude
\begin{align}
{\cal \hat{N}}_{i \lambda_1^{} \lambda_2^{} }^{} = \epsilon^{\mu_1^{}}_{}\bigl(q_1^{}, \lambda_1^{}\bigr) 
  \epsilon^{\mu_2^{}}_{}\bigl(q_2^{}, \lambda_2^{}\bigr) A_{i \mu_1^{} \mu_2^{}}^{}. \label{sub-amplitudes-gghh}
\end{align}
Then, the amplitude eq.~(\ref{eq:amp-gghh}) has a simpler form
\begin{align}
\bigl( {\cal M}_{gg}^{HH} \bigr)_{\lambda_1^{} \lambda_2^{}}^{}
 = \frac{G_F^{}\alpha_s^{} \hat{s} }{2\sqrt{2}\pi} \delta_{b_1^{} b_2^{}}^{}
\biggl\{
\Bigl[
3 m_H^2 \lambda_h^{}  D_H^{} \bigl( q_1^{} + q_2^{} \bigr) F_{\triangle}^{} + F_{\square} \Bigr] {\cal \hat{N}}_{1 \lambda_1^{} \lambda_2^{} }^{} + 
G_{\square}^{} {\cal \hat{N}}_{2 \lambda_1^{} \lambda_2^{} }^{}
\biggr\}. \label{amplitudes-gghh-2}
\end{align}
After a simple manipulation in the VBF frame, we find 
\begin{subequations}\label{eq:amp-parity-inv}
\begin{align}
{\cal \hat{N}}_{1 \lambda_1^{} \lambda_2^{} }^{} & = - \delta_{ \lambda_1^{}, \lambda_2^{} }^{}, \\
{\cal \hat{N}}_{2 \lambda_1^{} \lambda_2^{} }^{} & = - \delta_{ \lambda_1^{}, -\lambda_2^{} }^{}.
\end{align}
\end{subequations}
The parity invariance of the amplitude is apparent, $( {\cal M}_{gg}^{HH} )_{++}^{} = ( {\cal M}_{gg}^{HH} )_{--}^{}$ and $( {\cal M}_{gg}^{HH} )_{+-}^{} = ( {\cal M}_{gg}^{HH} )_{-+}^{}$. 
Up to now we have assumed the Higgs sector of the standard model. In this paper, we study the impact of parity symmetry violation in the GF process $gg\to HH$~\footnote{The parity symmetry violation in the GF process indicates charge-conjugation and parity (CP) symmetry violation in the Higgs sector.}. We can read off parity violating phases from a parity-odd process $gg \to HA$, where $A$ is a parity-odd Higgs boson. 
By using the tensor $A_{i \mu_1^{} \mu_2^{}}^{}$ for the process $gg \to HA$~\cite{Plehn:1996wb}, we find the amplitude in the VBF frame:
\begin{subequations}\label{amp-cp-odd}
\begin{align}
{\cal \hat{N}}_{1 \lambda_1^{} \lambda_2^{} }^{\mathrm{P\mathchar`-odd}} & = - i \lambda_1^{} \delta_{ \lambda_1^{}, \lambda_2^{} }^{}, \\
{\cal \hat{N}}_{2 \lambda_1^{} \lambda_2^{} }^{\mathrm{P\mathchar`-odd}} & = - i \lambda_1^{} \delta_{ \lambda_1^{}, -\lambda_2^{} }^{}.
\end{align}
\end{subequations}
In order to evaluate the impact of parity violation, we introduce two angles (or phases) $\xi_{1,2}^{}$ ($-\pi/2 \le \xi_{1,2}^{} \le \pi/2$) and write the amplitude as
\begin{subequations}\label{eq:cpv-phases}
\begin{align}
{\cal \hat{N}}_{1 \lambda_1^{} \lambda_2^{} }^{} & = - \delta_{ \lambda_1^{}, \lambda_2^{} }^{} e^{+i \lambda_1^{} \xi_1^{}}_{}, \\
{\cal \hat{N}}_{2 \lambda_1^{} \lambda_2^{} }^{} & = - \delta_{ \lambda_1^{}, -\lambda_2^{} }^{} e^{+i \lambda_1^{} \xi_2^{}}_{}.
\end{align}
\end{subequations}
The two phases $\xi_{1,2}^{}$ parametrise the magnitude of parity violation in the process $gg \to HH$ and independently affect the $\Delta \lambda = \lambda_1^{} - \lambda_2^{} = 0$ helicity states and $\Delta \lambda = \pm 2$ helicity states, respectively. 
We believe that this simplified introduction of parity violating phases is enough for studying the impact of parity violation on the azimuthal angle correlations.

\section{Azimuthal angle correlations}\label{sec:correlation}

\begin{table}[t]
\centering
{\renewcommand\arraystretch{1.5}
\begin{tabular}{|c|c||c|c|}
\hline
\multicolumn{2}{|c||}{ $F_i^{}[qq]$ } &
\multicolumn{2}{|c|}{  $F_i^{}[qg]$ } \\ 
\hline 
$F_0^{}[qq]$ & 
{\large 
$ \frac{1}{4}\ 
  \frac{ 1 + \cos^2_{}{\theta_1^{}} }{ \cos^2_{}{\theta_1^{}} }\ 
  \frac{ 1 + \cos^2_{}{\theta_2^{}} }{ \cos^2_{}{\theta_2^{}} } $
}
&
$F_0^{}[qg]$ & 
{\large 
$ \frac{1}{4}\ 
  \frac{ 1 + \cos^2_{}{\theta_1^{}} }{ \cos^2_{}{\theta_1^{}} }\ 
  \frac{ (1 + 3\cos^2_{}{\theta_2^{}})^2_{} }{ \sin^2_{}{\theta_2^{}} \cos^2_{}{\theta_2^{}} } $
} 
\\
$F_1^{}[qq]$ & 
{\large 
$ \frac{1}{4}\ 
  \frac{ 1 - \cos^2_{}{\theta_1^{}} }{ \cos^2_{}{\theta_1^{}} }\ 
  \frac{ 1 + \cos^2_{}{\theta_2^{}} }{ \cos^2_{}{\theta_2^{}} } $
}
&
$F_1^{}[qg]$ & 
{\large 
$ \frac{1}{4}\ 
  \frac{ 1 - \cos^2_{}{\theta_1^{}} }{ \cos^2_{}{\theta_1^{}} }\ 
  \frac{ (1 + 3\cos^2_{}{\theta_2^{}})^2_{} }{ \sin^2_{}{\theta_2^{}} \cos^2_{}{\theta_2^{}} } $
}
\\
$F_2^{}[qq]$ & 
{\large 
$ \frac{1}{4}\ 
  \frac{ 1 + \cos^2_{}{\theta_1^{}} }{ \cos^2_{}{\theta_1^{}} }\ 
  \frac{ 1 - \cos^2_{}{\theta_2^{}} }{ \cos^2_{}{\theta_2^{}} } $
}
& 
$F_2^{}[qg]$ &
{\large 
$ \frac{1}{4}\ 
  \frac{ 1 + \cos^2_{}{\theta_1^{}} }{ \cos^2_{}{\theta_1^{}} }\ 
  \frac{ 1 - \cos^2_{}{\theta_2^{}} }{ \cos^2_{}{\theta_2^{}} } $
}
\\
$F_3^{}[qq]$ & 
{\large 
$ \frac{1}{4}\ 
  \frac{ 1 - \cos^2_{}{\theta_1^{}} }{ \cos^2_{}{\theta_1^{}} }\ 
  \frac{ 1 - \cos^2_{}{\theta_2^{}} }{ \cos^2_{}{\theta_2^{}} } $
}
& 
$F_3^{}[qg]$ & 
{\large 
$ \frac{1}{4}\ 
  \frac{ 1 - \cos^2_{}{\theta_1^{}} }{ \cos^2_{}{\theta_1^{}} }\ 
  \frac{ 1 - \cos^2_{}{\theta_2^{}} }{ \cos^2_{}{\theta_2^{}} } $
}
\\
\hline
\hline
\multicolumn{2}{|c||}{ $F_i^{}[gq]$ } &
\multicolumn{2}{|c|}{  $F_i^{}[gg]$ } \\ 
\hline 
$F_0^{}[gq]$ & 
{\large 
$ \frac{1}{4}\ 
  \frac{ (1 + 3\cos^2_{}{\theta_1^{}})^2_{} }{ \sin^2_{}{\theta_1^{}} \cos^2_{}{\theta_1^{}} }\ 
  \frac{ 1 + \cos^2_{}{\theta_2^{}} }{ \cos^2_{}{\theta_2^{}} } $
}
&
$F_0^{}[gg]$ & 
{\large 
$ \frac{1}{4}\
  \frac{ (1 + 3\cos^2_{}{\theta_1^{}})^2_{} }{ \sin^2_{}{\theta_1^{}} \cos^2_{}{\theta_1^{}} }\ 
  \frac{ (1 + 3\cos^2_{}{\theta_2^{}})^2_{} }{ \sin^2_{}{\theta_2^{}} \cos^2_{}{\theta_2^{}} } $
} 
\\
$F_1^{}[gq]$ & 
{\large 
$ \frac{1}{4}\ 
  \frac{ 1 - \cos^2_{}{\theta_1^{}} }{ \cos^2_{}{\theta_1^{}} }\ 
  \frac{ 1 + \cos^2_{}{\theta_2^{}} }{ \cos^2_{}{\theta_2^{}} } $
}
&
$F_1^{}[gg]$ & 
{\large 
$ \frac{1}{4}\
  \frac{ 1 - \cos^2_{}{\theta_1^{}} }{ \cos^2_{}{\theta_1^{}} }\ 
  \frac{ (1 + 3\cos^2_{}{\theta_2^{}})^2_{} }{ \sin^2_{}{\theta_2^{}} \cos^2_{}{\theta_2^{}} } $
} 
\\
$F_2^{}[gq]$ & 
{\large 
$ \frac{1}{4}\ 
  \frac{ (1 + 3\cos^2_{}{\theta_1^{}})^2_{} }{ \sin^2_{}{\theta_1^{}} \cos^2_{}{\theta_1^{}} }\ 
  \frac{ 1 - \cos^2_{}{\theta_2^{}} }{ \cos^2_{}{\theta_2^{}} } $
}
&
$F_2^{}[gg]$ & 
{\large 
$ \frac{1}{4}\
  \frac{ (1 + 3\cos^2_{}{\theta_1^{}})^2_{} }{ \sin^2_{}{\theta_1^{}} \cos^2_{}{\theta_1^{}} }\ 
  \frac{ 1 - \cos^2_{}{\theta_2^{}} }{ \cos^2_{}{\theta_2^{}} } $
} 
\\
$F_3^{}[gq]$ & 
{\large 
$ \frac{1}{4}\ 
  \frac{ 1 - \cos^2_{}{\theta_1^{}} }{ \cos^2_{}{\theta_1^{}} }\ 
  \frac{ 1 - \cos^2_{}{\theta_2^{}} }{ \cos^2_{}{\theta_2^{}} } $
}
& 
$F_3^{}[gg]$ & 
{\large 
$ \frac{1}{4}\ 
  \frac{ 1 - \cos^2_{}{\theta_1^{}} }{ \cos^2_{}{\theta_1^{}} }\ 
  \frac{ 1 - \cos^2_{}{\theta_2^{}} }{ \cos^2_{}{\theta_2^{}} } $
}
\\
\hline
\end{tabular}
\caption{\small Functions $F_i^{}[a_1^{}a_2^{}]$ in eq.~(\ref{full-amp-squared-ggh}) for the $qq$ initiated sub-process $(a_1^{}, a_2^{})=(q, q)$ (upper-left), the $qg$ initiated sub-process $(a_1^{}, a_2^{})=(q, g)$ (upper-right), the $gq$ initiated sub-process $(a_1^{}, a_2^{})=(g, q)$ (lower-left) and the $gg$ initiated sub-process $(a_1^{}, a_2^{})=(g, g)$ (lower-right). $\theta_{1,2}^{}$ are defined in the $q_{1,2}^{}$ Breit frames eqs.~(\ref{eq:breit1}) and (\ref{eq:breit2}.}
\label{table:splitting-functions}
}
\end{table}

In this section we present a detailed study of the azimuthal angle correlations of the two jets by using the helicity amplitudes presented in Section~\ref{sec:Analytic_formula}. 

\subsection{The gluon fusion process}\label{sec:gf-process}

The azimuthal angle correlations of the two jets can be analytically apparent, once we obtain the squared VBF amplitude. There are four gluon fusion (GF) sub-processes $(V_1^{}, V_2^{})=(g, g)$, namely the $qq$ initiated sub-process $(a_1^{}, a_2^{})=(q, q)$, the $qg$ initiated sub-processes $(a_1^{}, a_2^{})=(q, g), (g, q)$ and the $gg$ initiated sub-process $(a_1^{}, a_2^{})=(g, g)$. 
The squared VBF amplitude for the four GF sub-processes has the following compact form, after averaging over the initial state colours and helicities and summing over the final state colours and helicities:
\begin{align}
\overline{\sum_{\mathrm{col.}}}\ \overline{\sum_{\sigma_i^{}=\pm}} \bigl| {\cal M}_{\sigma_1^{} \sigma_2^{}}^{\sigma_3^{} \sigma_4^{}} \bigr|^2_{}  =
\frac{( 4\pi \alpha_s)^2_{} C_{a_1^{}a_2^{}}^{} }{ Q_1^2 Q_2^2 } 
\overline{\sum_{b_1^{},b_2^{}}}
\biggl\{
& F_0^{}[a_1^{}a_2^{}] 
\Bigl(
|\hat{\cal M}_{++}^{}|^2_{} + |\hat{\cal M}_{+-}^{}|^2_{} + |\hat{\cal M}_{-+}^{}|^2_{} + |\hat{\cal M}_{--}^{}|^2_{} 
\Bigr) \nonumber \\
& -
2 F_1^{}[a_1^{}a_2^{}] 
\Bigl[
{\it Re}\bigl( \hat{\cal M}_{++}^{} \hat{\cal M}_{-+}^* \bigr) +
{\it Re}\bigl( \hat{\cal M}_{+-}^{} \hat{\cal M}_{--}^* \bigr)
\Bigr] \cos{2\phi_1^{}} \nonumber \\
& -
2 F_2^{}[a_1^{}a_2^{}] 
\Bigl[
{\it Re}\bigl( \hat{\cal M}_{+-}^{} \hat{\cal M}_{++}^* \bigr) +
{\it Re}\bigl( \hat{\cal M}_{--}^{} \hat{\cal M}_{-+}^* \bigr)
\Bigr] \cos{2\phi_2^{}} \nonumber \\
& + 
2 F_3^{}[a_1^{}a_2^{}] 
{\it Re}\bigl( \hat{\cal M}_{++}^{} \hat{\cal M}_{--}^* \bigr) 
\cos{2(\phi_1^{}-\phi_2^{})} \nonumber \\
& + 
2 F_3^{}[a_1^{}a_2^{}] 
{\it Re}\bigl( \hat{\cal M}_{+-}^{} \hat{\cal M}_{-+}^* \bigr) 
\cos{2(\phi_1^{}+\phi_2^{})} \nonumber \\
& + 
\bigl( {\it Re} \to {\it Im}, \cos{} \to \sin{} \bigr)
\biggr\}.\label{full-amp-squared-ggh}
\end{align}
In order to simplify our writing, we introduce a notation $\hat{\cal M}_{\lambda_1^{} \lambda_2^{}}^{}$ which denotes the helicity amplitude $( {\cal M}_{gg}^{HH} )_{\lambda_1^{} \lambda_2^{}}^{}$. $C_{a_1^{} a_2^{}}^{}$ are the colour factors from the splitting processes $a_1^{} \to a_{3}^{} + g_1^*$ and $a_2^{} \to a_{4}^{} + g_2^*$, and they take values $C_{qq}^{} = 16/9$, $C_{qg}^{} = C_{gq}^{} = 4$ and $C_{gg}^{}=9$. $F_i^{}[a_1^{}a_2^{}]$ are functions of the kinematic variables $\theta_{1,2}^{}$ defined in the $q_{1,2}^{}$ Breit frames eqs.~(\ref{eq:breit1}) and (\ref{eq:breit2}), and summarised in Table~\ref{table:splitting-functions}. The azimuthal angles $\phi_{1,2}^{}$ are also defined in the Breit frames. $b_{1,2}^{}$ are the colour indices of the intermediate gluons and an average for $b_{1,2}^{}$ is performed, see eq.~(\ref{eq:amp-gghh}). The colour factors $C_{a_1^{} a_2^{}}^{}$ and the functions $F_i^{}[a_1^{}a_2^{}]$ for antiquarks are the same as those for quarks, since the QCD interaction does not distinguish quarks and antiquarks. \\

\begin{figure}
\centering
\includegraphics[scale=0.62]{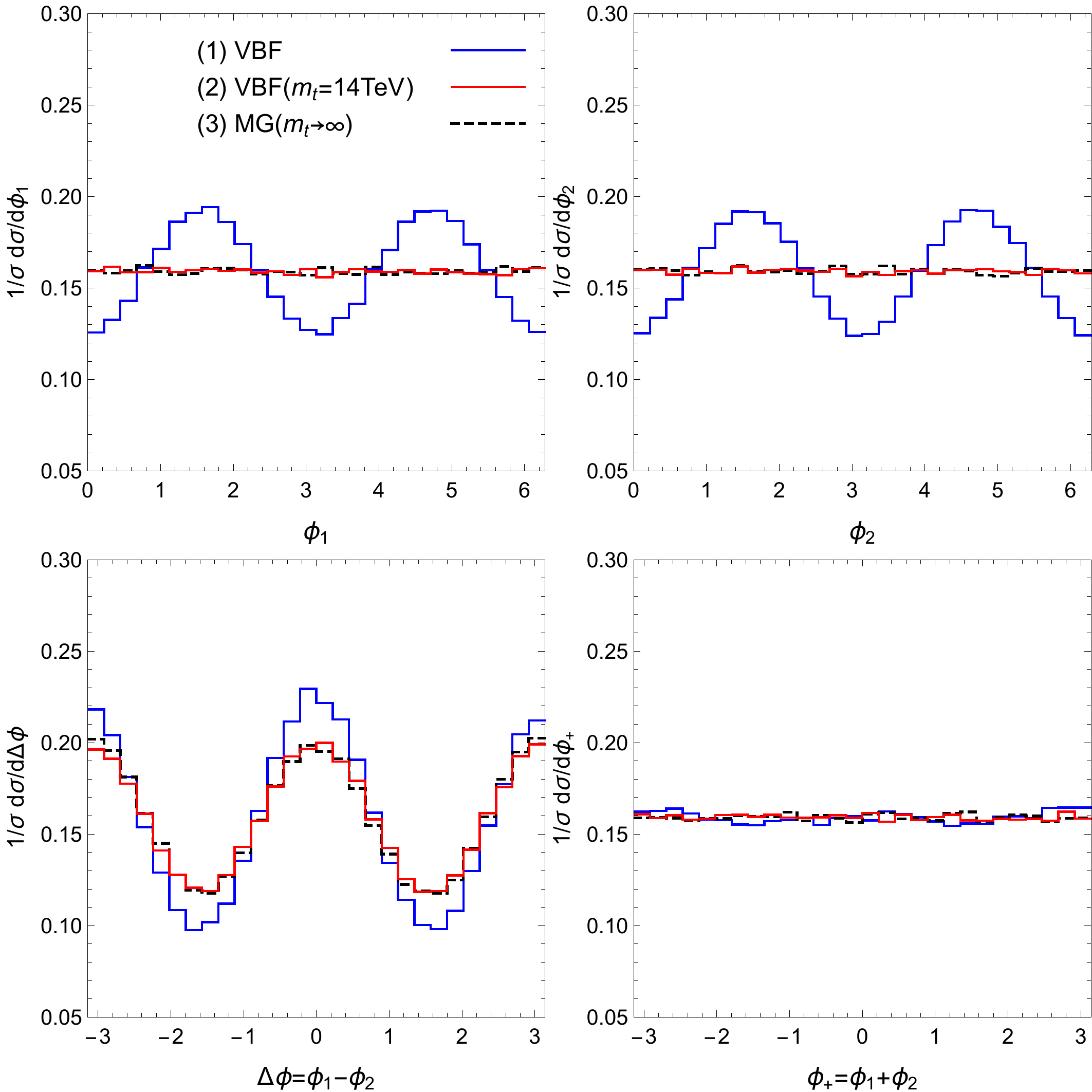}
\caption{\small 
The normalized differential cross section of the GF process as a function of $\phi_1^{}$ (upper left), $\phi_2^{}$ (upper right), $\Delta \phi$ (lower left) and $\phi_+^{}$ (lower left). The correspondence between curves and simulation methods is shown inside the upper left panel: (1) the blue curve represents the result according to our analytic cross section formula, to which only the VBF diagrams contribute, (2) the red curve represents the result according to our analytic cross section formula with $m_t^{}$ being $14$ TeV, (3) the black dashed curve represents the exact LO result with the effective interactions between gluons and the Higgs bosons (infinite $m_t^{}$ limit).}
\label{figure:phi-dist-1}
\end{figure}

The first term in the right hand side (RHS) of eq.~(\ref{full-amp-squared-ggh}) contributes to the inclusive cross section after a phase space integration, while the other terms give the azimuthal angle distributions of the two jets.  
The azimuthal angles $\phi_{1,2}^{}$ of the two jets are defined in the $q_{1,2}^{}$ Breit frames, respectively, and they remain the same in the VBF frame. In the limit that each of the two jets in the proton-proton (pp) frame is collinear to the incoming parton that emits it (collinear limit), the emitted two intermediate vector bosons also move on the $z$-axis. After rotating the two jets and the two Higgs bosons around the $z$-axis in such a way that the two Higgs bosons have zero azimuthal angle (Let us remind that the two Higgs bosons have zero azimuthal angle in the VBF frame, see eq.~(\ref{eq:VBF-four-momenta})) and after a single boost along the $z$-axis, all of these particles can be studied in the VBF frame.
Therefore, in the collinear limit, the azimuthal angles of the two jets after the single rotation around the $z$-axis are identical to $\phi_1^{}$ and $\phi_2^{}$. We apply the VBF cuts and an upper transverse momentum $p_T^{}$ cut on the jets in the pp frame and these cuts reproduce the collinear limit to some extent. 
Hence the azimuthal angles of the two jets in the pp frame after the rotation around the $z$-axis should not be very different from $\phi_{1,2}^{}$ defined in the $q_{1,2}^{}$ Breit frames.
We perform the rotation of the two jets and the two Higgs bosons around the $z$-axis in the following way:
\begin{enumerate}
\item Go to the centre-of-mass (c.m.) frame of the two Higgs bosons and then rotate the two Higgs bosons around the $z$-axis by $\tilde{\phi}$ in a way that the two Higgs bosons have zero azimuthal angle.
\item Rotate the two jets in the pp frame around the $z$-axis by $\tilde{\phi}$.
\item Measure the azimuthal angles of the two jets.
\end{enumerate}
In the collinear limit, it is clear that the azimuthal angles of the two jets measured after this rotation coincide with $\phi_{1,2}^{}$.
Note that this rotation is necessary for the azimuthal angles and the sum of them to show meaningful distributions, because the process $pp \to HHjj$ in the pp frame is completely symmetric around the $z$-axis. This rotation is, however, not needed for the difference of the two azimuthal angles. 
Before we show numerical results, we point out characteristic features of the GF sub-process in the standard model (SM) already expected from the analytic formula eq.~(\ref{full-amp-squared-ggh}):
\begin{itemize}
\item The azimuthal angles of the two jets show the same distribution due to the parity invariance of the amplitude $\hat{\cal M}_{\lambda_1^{} \lambda_2^{}}^{}$, see eq.~(\ref{eq:amp-parity-inv}).
\item All of the azimuthal angle observables show cosine distributions, again due to the parity invariance of the amplitude.
\end{itemize}
Violation of the parity invariance of the amplitude should appear as a deviation from the above expectations. This case will be studied at the end of this subsection. \\

We show numerical results for the $14$ TeV LHC. We do not study decays of the two Higgs bosons and assume that they can be reconstructed. An outgoing quark, antiquark or gluon is identified as a jet. 
The following set of parameters are chosen: $m_H^{}=125.5$ GeV, $m_t^{}=173.5$ GeV and $\alpha_s^{}(m_Z^{})=0.13$. We use the CTEQ6L1~\cite{Pumplin:2002vw} set for the parton distribution functions (PDFs) and have chosen the input value for the strong coupling constant accordingly.
For the scales in the PDFs, we choose a fixed value of $25$ GeV, which corresponds to the lower cutoff on the transverse momentum $p_T^{}$ of the jets (see below). 
The scales in the strong couplings are chosen as $\alpha_s^{}(\sqrt{\hat{s}})^2_{} \alpha_s^{}(150 \mathrm{GeV})^2_{}$, where $\sqrt{\hat{s}}$ is the invariant mass of the two Higgs bosons. Using the two different scales in the strong couplings can be considered as a better choice, since we look at only a kinematic region where the virtualities of the gluons are small ($Q_{1,2}^{}\to 0$) and this separates the two splitting processes from the process $gg\to HH$ in time-scale. 
The scales in the strong couplings of the splitting processes correspond to the upper cutoff on the $p_T^{}$ of the jets.
The following cuts are applied on the rapidity $y$ and $p_T^{}$ of the two jets in the pp frame:
\begin{align}
-5 < y_2^{} < 0 < y_1^{} < 5, \ \ \ \ \
y_1^{} - y_2^{} > 5, \ \ \ \ \
25\ \mathrm{GeV} < p_T^{} < 150\ \mathrm{GeV}. \label{eq:kin-region}
\end{align}
The above rapidity cuts are the VBF cuts.
As we have already mentioned in Section~\ref{sec:intro}, the VBF cuts and the upper $p_T^{}$ cut enhance the contributions from the VBF sub-processes. This can be understood as follows. The virtuality $Q_1^{}$ of the intermediate vector boson $V_1^{}$ is 
\begin{align}
Q_1^2 &= - \bigl(k_1^{}-k_{3}^{}\bigr)^2_{} \nonumber \\
&= 2 E_1^{} E_{3}^{} \bigl( 1 - \cos{\theta_{13}^{}} \bigr) \nonumber \\
&= E_1^{} E_{3}^{} \theta_{13}^2 + O( \theta_{13}^4 ),
\end{align}
where the momentum assignment given in eq.~(\ref{eq:mom-assign}) is used and $\theta_{13}^{}$ is the angle between $\vec{k}_1^{}$ and $\vec{k}_{3}^{}$. The VBF cuts make $\vec{k}_{3}^{}$ collinear to $\vec{k}_1^{}$ ($\theta_{13}^{}\to0$), then $Q_1^{}$ is decreased and the VBF amplitude is enhanced. The upper $p_T^{}$ cut additionally enhances the VBF amplitude. In a collinear case ($\theta_{13}^{}\to0$), the $p_T^{}$ of $k_3^{}$ is
\begin{align}
p_T^2 &=  E_{3}^2 \sin^2{\theta_{13}^{}} \nonumber \\
      &=  E_{3}^2 \theta_{13}^2 + O( \theta_{13}^4 ),
\end{align}
so the upper $p_T^{}$ cut reasonably implies the upper cut on $Q_1^{}$.
Only the VBF cuts may be enough to enhance contributions from the VBF sub-processes. 
However, we need to impose the upper $p_T^{}$ cut, too, because we perform the two approximations in calculating the VBF amplitude for the GF sub-process and these approximations are justified only when $Q_1^2 < \hat{s}$ and $Q_2^2 < \hat{s}$. 
Throughout our analyses, the four-momentum of the jet which has a positive rapidity $y_1^{}$ in the pp frame is used for calculating its azimuthal angle labelled as $\phi_1^{}$ and that of the other jet which has a negative rapidity $y_2^{}$ in the pp frame is used for calculating its azimuthal angle labelled as $\phi_2^{}$. Azimuthal angles labelled as $\phi_{1,2}^{}$ in our numerical results shown below are not those defined in the $q_{1,2}^{}$ Breit frames anymore. 
The phase space integration and event generations are performed with the programs {\tt BASES} and {\tt SPRING}~\cite{Kawabata:1995th}. The scalar loop functions are calculated with the program {\tt FF}~\cite{vanOldenborgh:1989wn}.\\

\begin{figure}[t]
\centering
\includegraphics[scale=0.5]{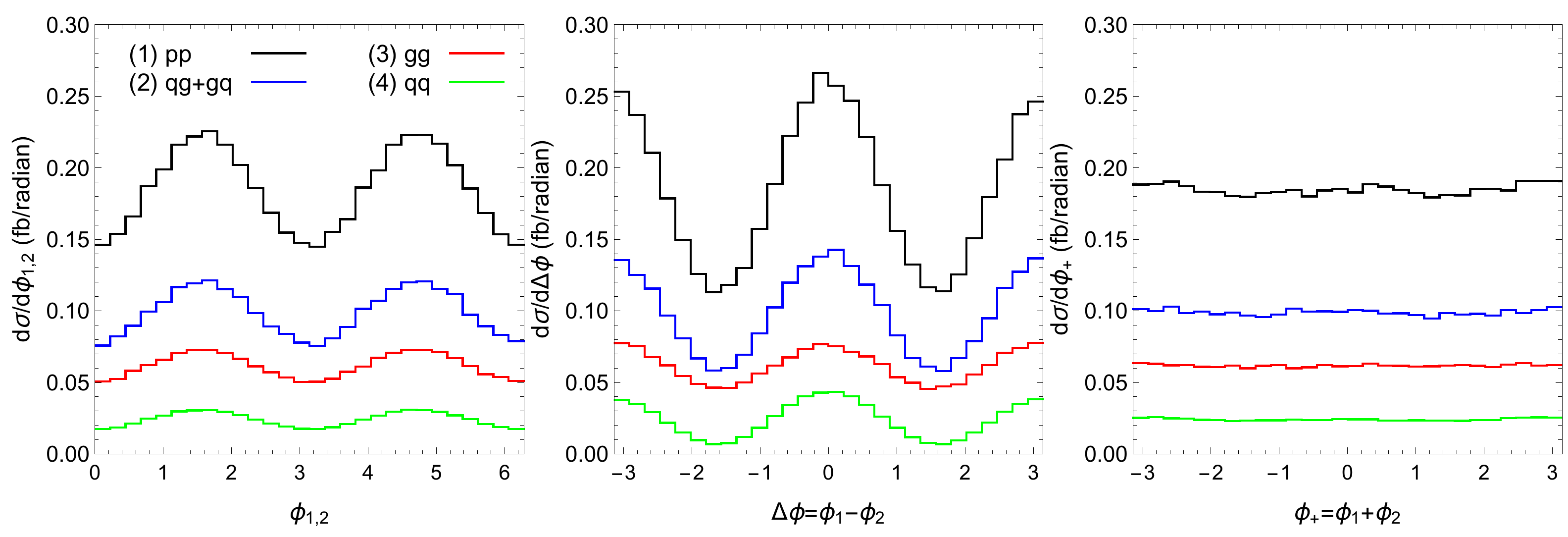}
\caption{\small 
The differential cross section in unit of femto barn as a function of $\phi_{1,2}^{}$ (left), $\Delta \phi$ (middle) and $\phi_+^{}$ (right), contributed by different GF sub-processes. 
All of the sub-processes contribute to the black curve,  the $qg$ and $gq$ initiated sub-processes to the blue curve, the $gg$ initiated sub-process to the red curve and the $qq$ initiated sub-process to the green curve.}
\label{figure:phi-crosssec}
\end{figure}

\begin{table}[t]
\centering
\begin{tabular}{|c|ccc|c|}
\hline
 & $\sigma_{HHqq}^{}$ (fb) & $\sigma_{HHgg}^{}$ (fb) & $\sigma_{HHqg}^{}$ (fb) & $\sigma_{HHjj}^{}$ (fb) \\
\hline
VBF & $0.1514(8)$ & $0.3876(2)$ & $0.6215(3)$ & $1.1607(6)$ \\
\hline 
VBF ($m_t^{}=14$TeV) & $0.2647(1)$ & $0.6108(3)$ & $1.0743(5)$ & $1.9495(9)$ \\
\hline
MG ($m_t^{} \to \infty$) & $0.2788(3)$ & $0.6289(6)$ & $1.112(1)$ & $2.019(1)$ \\
\hline
\end{tabular}
\caption{\small
The inclusive cross sections in unit of femto barn for various final states in pp collisions, produced by three different methods. VBF: our analytic cross section formula, to which only the VBF diagrams contribute. VBF ($m_t^{}=14$TeV): our analytic cross section formula with $m_t^{}$ being 14 TeV. MG ($m_t^{} \to \infty$): the exact LO amplitude with the effective interactions between gluons and the Higgs bosons. The statistical uncertainty for the last digit is shown in the parenthesis.}
\label{table:inclusive-crosssection}
\end{table}

In Figure~\ref{figure:phi-dist-1} we show the normalised differential cross section as a function of $\phi_1^{}$ (upper left), $\phi_2^{}$ (upper right), $\Delta \phi = \phi_1^{} - \phi_2^{}$ (lower left) and $\phi_+^{} = \phi_1^{} + \phi_2^{}$ (lower right). 
The blue solid curve, labelled as (1) VBF, represents the result according to our analytic cross section formula, to which only the VBF diagrams contribute. 
The scalar loop functions in the form factors are calculated by using the finite $m_t^{}$ value.
The red solid curve, labelled as (2) VBF($m_t^{}=14$TeV), represents the result according to our analytic cross section formula with $m_t^{}$ in the form factors being an extremely large value, $m_t^{}=14$ TeV.
Finally the black dashed curve, labelled as (3) MG($m_t^{} \to \infty$), represents the result according to the exact leading order (LO) amplitude with effective interactions between gluons and the Higgs bosons (infinite $m_t^{}$ limit). The event generation for the result (3) is performed by implementing the following effective Lagrangian density~\cite{Shifman:1979eb} into an UFO file~\cite{Degrande:2011ua} with the help of {\tt FeynRule}~\cite{Christensen:2008py} version 1.6.18 and subsequently using the UFO file in {\tt MadGraph5\verb|_|aMC@NLO}\cite{Alwall:2014hca} version 5.2.2.1:
\begin{align}
{\cal L} = \frac{\alpha_s^{}}{12\pi} \bigl(\sqrt{2}G_F^{}\bigr)^{1/2}_{} F_{\mu\nu}^a F_{}^{a, \mu\nu}H - \frac{\alpha_s^{}}{24\pi} \sqrt{2}G_F^{} F_{\mu\nu}^a F_{}^{a, \mu\nu}HH,
\end{align}
where $F_{\mu\nu}^a$ is the gluon field strength tensor and $H$ is the Higgs boson field. The interactions between gluons and the Higgs bosons in this approach can be considered as those induced by a quark loop where the mass of the quark is infinitely large. 
The consistency between the result (2) and the result (3) in all of the panels shows that our analytic cross section formula is a good approximation to the exact cross section formula. 
The discrepancy between the result (1) and the result (2) in the $\phi_{1,2}^{}$ plots particularly shows the importance of using a finite $m_t^{}$ value in the azimuthal angle distributions. 
The reason why we do not find correlated distributions in $\phi_{1,2}^{}$ in the large $m_t^{}$ results can be understood from the squared VBF amplitude in eq.~(\ref{full-amp-squared-ggh}). 
The second and third terms in the RHS shows that the correlations in $\phi_{1,2}^{}$ arise from the interference of the amplitudes $\hat{\cal M}_{\lambda_1^{} \lambda_2^{}}^{}$ for $\Delta \lambda = 0$ and $\Delta \lambda = \pm 2$ states, where $\Delta \lambda = \lambda_1^{}-\lambda_2^{}$. 
In the large $m_t^{}$ limit, the amplitude for $\Delta \lambda = \pm 2$ states vanishes and hence the correlations also vanish.
The GF sub-process exhibits the largest correlation in $\Delta \phi$ and an almost zero correlation in $\phi_+^{}$. The squared VBF amplitude in eq.~(\ref{full-amp-squared-ggh}) tells us that the correlation in $\Delta \phi$ arises from the interference of the amplitudes $\hat{\cal M}_{\lambda_1^{} \lambda_2^{}}^{}$ for $\Delta \lambda = 0$ states (the fourth term in the RHS) and the correlation in $\phi_+^{}$ arises from the interference of the amplitudes for $\Delta \lambda = \pm 2$ states (the fifth term in the RHS). The reason of the large correlation in $\Delta \phi$ and the small correlation in $\phi_+^{}$ is because the amplitudes for $\Delta \lambda = 0$ states are much larger than those for $\Delta \lambda = \pm 2$ states in a large part of the phase space (this will be confirmed explicitly below soon). 
We briefly mention differences between the processes $pp\to HHjj$ and $pp\to Hjj$  in the correlations. The process $gg\to H$ has non-zero amplitudes $\hat{\cal M}_{\lambda_1^{} \lambda_2^{}}^{}$ only for $\Delta \lambda = 0$ states. Therefore, the process $pp\to Hjj$ exhibits a large correlation in $\Delta \phi$~\cite{Plehn:2001nj, Hankele:2006ma, Klamke:2007cu, Hagiwara:2009wt} but no correlations in $\phi_{1,2}^{}$ and $\phi_+^{}$. \\

\begin{figure}[t]
\centering
\includegraphics[scale=0.5]{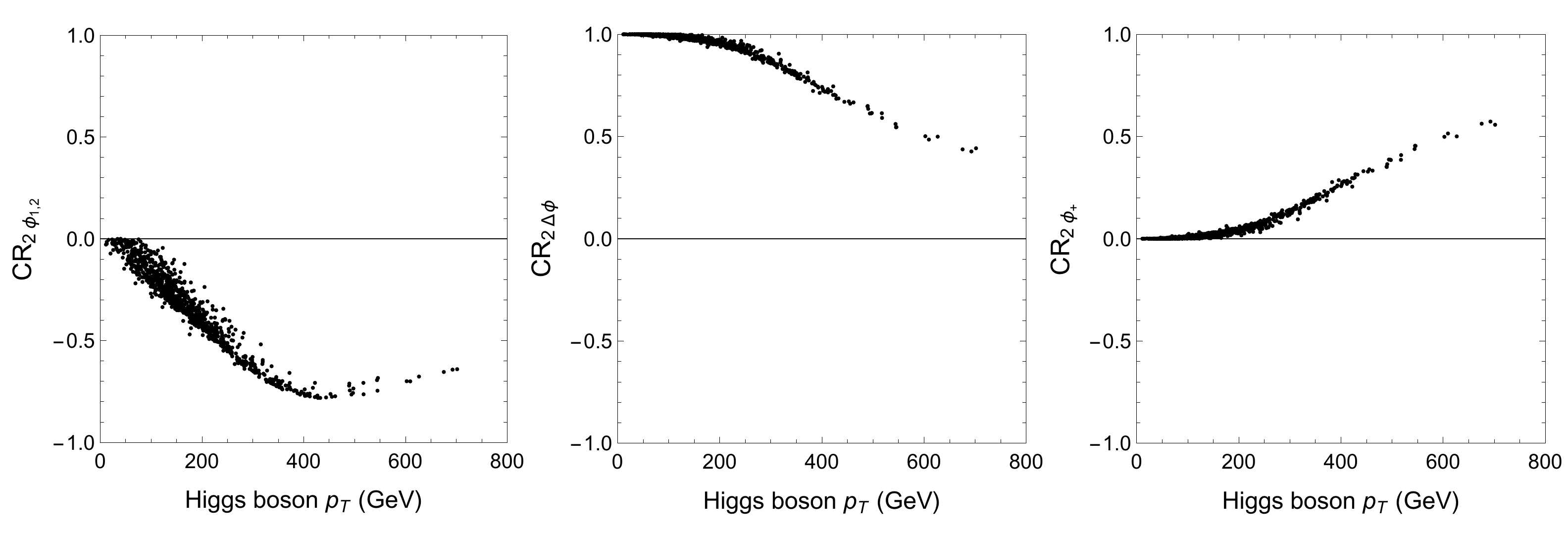}
\caption{\small List plots of $CR_{2\phi_{1,2}^{}}^{}$ (left), $CR_{2\Delta \phi}^{}$ (middle) and $CR_{2\phi_+^{}}^{}$ (right) defined in eq.~(\ref{eq:corr-functions}) with the $p_T^{}$ of the Higgs boson (in GeV) in the c.m. frame of the two Higgs bosons.}
\label{figure:corr-1}
\end{figure}

In Figure~\ref{figure:phi-crosssec} we show the differential cross section as a function of $\phi_{1,2}^{}$ (left), $\Delta \phi = \phi_1^{} - \phi_2^{}$ (middle) and $\phi_+^{} = \phi_1^{} + \phi_2^{}$ (right), contributed by all of the sub-processes (black curve, labelled as (1) $pp$), by the $qg$ and $gq$ initiated sub-processes (blue curve, labelled as (2) $qg+gq$), by the $gg$ initiated sub-process (red curve, labelled as (3) gg) and by the $qq$ initiated sub-process (green curve, labelled as (4) qq). All of the numerical results hereafter in this subsection are produced by using our analytic cross section formula. 
As we have already discussed before showing the numerical results and have actually confirmed in Figure~\ref{figure:phi-dist-1}, $\phi_1^{}$ and $\phi_2^{}$ show the same distribution due to the parity invariance of the amplitude in the SM. Therefore, we show only the $\phi_1^{}$ distribution and label it $\phi_{1,2}^{}$ instead of showing the both distributions, until we study parity violation. In Table~\ref{table:inclusive-crosssection}, we present the inclusive cross sections of $qqHH$, $ggHH$, $qgHH$ final states and the sum of these final states in pp collisions. The calculation methods are the same as those used in Figure~\ref{figure:phi-dist-1}. A good agreement (within 5\%) between the second row and the third row for each cross section again confirms the validity of our analytic cross section formula. A large discrepancy between the first row and the second (or third) row for each cross section can be considered as another important result of using the finite $m_t^{}$ value. \\

\begin{figure}
\centering
\includegraphics[scale=0.5]{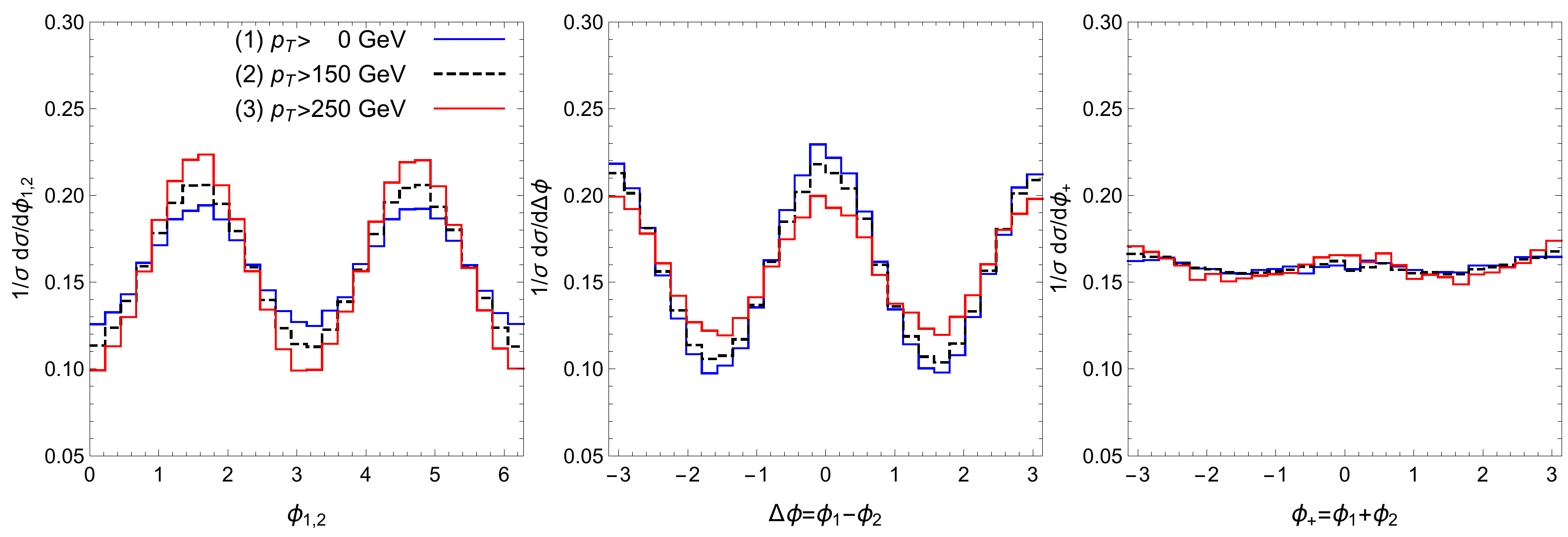}
\caption{\small 
The normalised differential cross section of the GF process as a function of $\phi_{1,2}^{}$ (left), $\Delta \phi$ (middle) and $\phi_+^{}$ (right), with three different values for lower cutoff on $p_T^{}$ of the Higgs boson in the c.m. frame of the two Higgs bosons. The correspondence between the curves and the cutoff values is shown inside the left panel.}
\label{figure:phi-dist-ptcut}
\end{figure}

In order to study how the azimuthal angle correlations depend on the kinematics of the two Higgs bosons, we introduce the following three quantities:
\begin{subequations}\label{eq:corr-functions}
\begin{align}
CR_{2\phi_{1,2}^{}}^{} = -
2
\frac{
{\it Re}\bigl( \hat{\cal M}_{++}^{} \hat{\cal M}_{-+}^* \bigr) +
{\it Re}\bigl( \hat{\cal M}_{+-}^{} \hat{\cal M}_{--}^* \bigr)
}
{
|\hat{\cal M}_{++}^{}|^2_{} + |\hat{\cal M}_{+-}^{}|^2_{} + |\hat{\cal M}_{-+}^{}|^2_{} + |\hat{\cal M}_{--}^{}|^2_{} 
}, \\
CR_{2\Delta \phi}^{} = 
+2
\frac{
{\it Re}\bigl( \hat{\cal M}_{++}^{} \hat{\cal M}_{--}^* \bigr) 
}
{
|\hat{\cal M}_{++}^{}|^2_{} + |\hat{\cal M}_{+-}^{}|^2_{} + |\hat{\cal M}_{-+}^{}|^2_{} + |\hat{\cal M}_{--}^{}|^2_{} 
}, \\
CR_{2\phi_+^{}}^{} = 
+2
\frac{
{\it Re}\bigl( \hat{\cal M}_{+-}^{} \hat{\cal M}_{-+}^* \bigr) 
}
{
|\hat{\cal M}_{++}^{}|^2_{} + |\hat{\cal M}_{+-}^{}|^2_{} + |\hat{\cal M}_{-+}^{}|^2_{} + |\hat{\cal M}_{--}^{}|^2_{} 
}.
\end{align}
\end{subequations}
These are simply the coefficients of the azimuthal angle dependent terms divided by the coefficient of the azimuthal angle independent term in eq.~(\ref{full-amp-squared-ggh}). 
Although the functions $F_i^{}[a_1^{}a_2^{}]$ are omitted, these quantities are useful in that an azimuthal angle observable should show a strong correlation in a kinematic region where the corresponding quantity $CR_i^{}$ is enhanced. 
We find that all of the quantities $CR_i^{}$ are well correlated with the $p_T^{}$ of the Higgs boson in the c.m. frame of the two Higgs bosons, as shown in Figure~\ref{figure:corr-1}. Figure~\ref{figure:corr-1} presents list plots of $CR_{2\phi_{1,2}^{}}^{}$ (left), $CR_{2\Delta \phi}^{}$ (middle) and $CR_{2\phi_+^{}}^{}$ (right) with the $p_T^{}$ of the Higgs boson in the c.m. frame of the two Higgs bosons. 
The large value of $CR_{2\Delta \phi}^{}$ and the small value of $CR_{2\phi_+^{}}^{}$ in the most part of the phase space explain the large correlation in $\Delta \phi$ and the small correlation in $\phi_+^{}$ observed in Figure~\ref{figure:phi-dist-1}. 
The plots indicate that the correlation in $\Delta \phi$ is enhanced as the $p_T^{}$ of the Higgs boson is decreased, while the correlations in $\phi_{1,2}^{}$ and $\phi_+^{}$ are enhanced as the $p_T^{}$ of the Higgs boson is increased. 
This is confirmed in Figure~\ref{figure:phi-dist-ptcut}. In Figure~\ref{figure:phi-dist-ptcut}, we show the normalised differential cross section as a function of $\phi_{1,2}^{}$ (left), $\Delta \phi = \phi_1^{} - \phi_2^{}$ (middle) and $\phi_+^{} = \phi_1^{} + \phi_2^{}$ (right) with different values for the lower cutoff on the $p_T^{}$ of the Higgs boson in the c.m. frame of the two Higgs bosons, blue solid curve: $p_T^{}>0$ GeV (no cut), black dashed curve: $p_T^{}>150$ GeV, and red solid curve: $p_T^{}>250$ GeV. \\

\begin{figure}[t]
\centering
\includegraphics[scale=0.5]{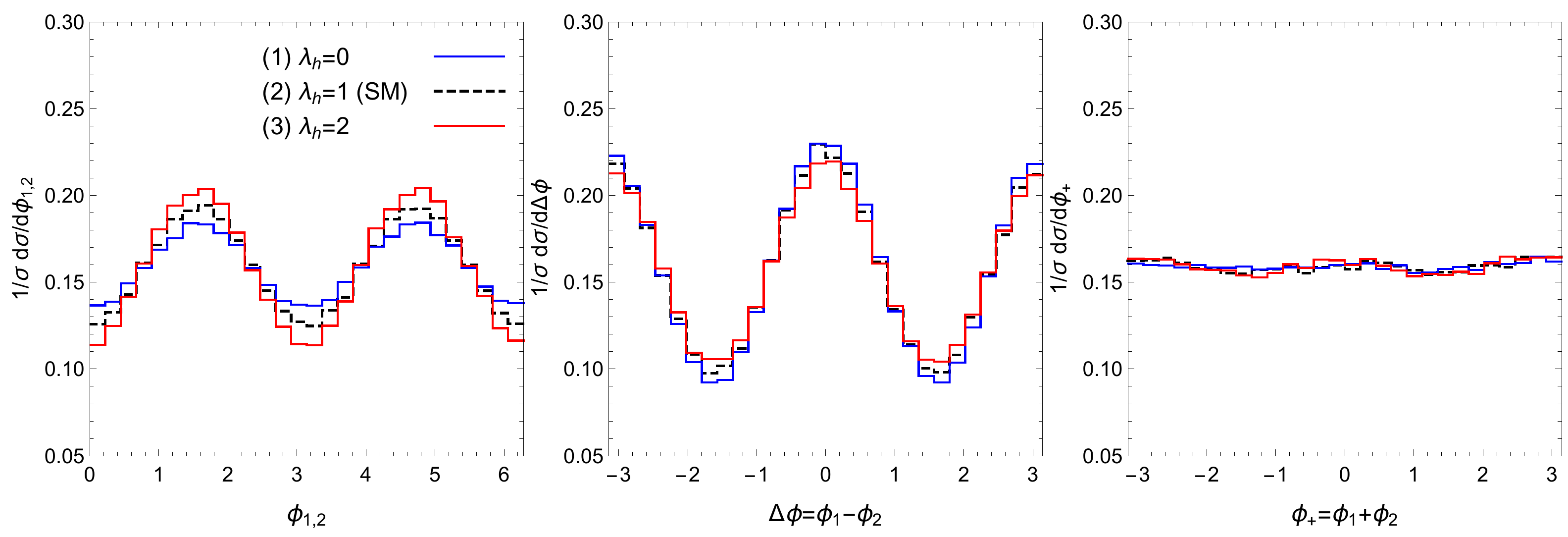}
\caption{\small 
The normalised differential cross section of the GF process as a function of $\phi_{1,2}^{}$ (left), $\Delta \phi$ (middle) and $\phi_+^{}$ (right), with three different values for the triple Higgs self-coupling re-scaling factor $\lambda_h^{}$. The correspondence between the curves and the values for $\lambda_h^{}$ is shown inside the left panel}
\label{figure:phi-dist-lambda}
\end{figure}

Next, we study how the azimuthal angle correlations depend on the triple Higgs self-coupling.  
Eqs.~(\ref{amplitudes-gghh-2}) and (\ref{eq:amp-parity-inv}) tell us that $\lambda_h^{}$, which is the factor re-scaling the triple Higgs self-coupling, affects only the amplitude $\hat{\cal M}_{\lambda_1^{} \lambda_2^{}}^{}$ for $\Delta \lambda = 0$ states. 
However, as is clear from the quantities in eq.~(\ref{eq:corr-functions}), $\lambda_h^{}$ affects all of the correlations. For an extreme example, if we could make the amplitude for $\Delta \lambda = 0$ states to be identically zero (this is actually not possible because the amplitude depend not only on $\lambda_h^{}$ but also on the kinematic of the process $gg\to HH$), we would observe a large correlation only in $\phi_+^{}$. 
In Figure~\ref{figure:phi-dist-lambda}, we study the correlations with three different values for $\lambda_h^{}$, the blue solid curve: $\lambda_h^{}=0$, the black dashed curve: $\lambda_h^{}=1$ (the SM prediction), and the red solid curve: $\lambda_h^{}=2$. 
The impact of a non-standard value for $\lambda_h^{}$ ($\lambda_h^{}\ne1$) in the distributions is visible but not large. Actually this is much smaller than that in other observables, such as the $p_T^{}$ of the Higgs boson~\cite{Dolan:2012rv,Baglio:2012np} or the invariant mass of the two Higgs bosons~\cite{Baglio:2012np,Barger:2013jfa}, of the inclusive process $pp\to HH$. The azimuthal angle correlations may not be useful to probe $\lambda_h^{}$. \\

\begin{figure}
\centering
\includegraphics[scale=0.5]{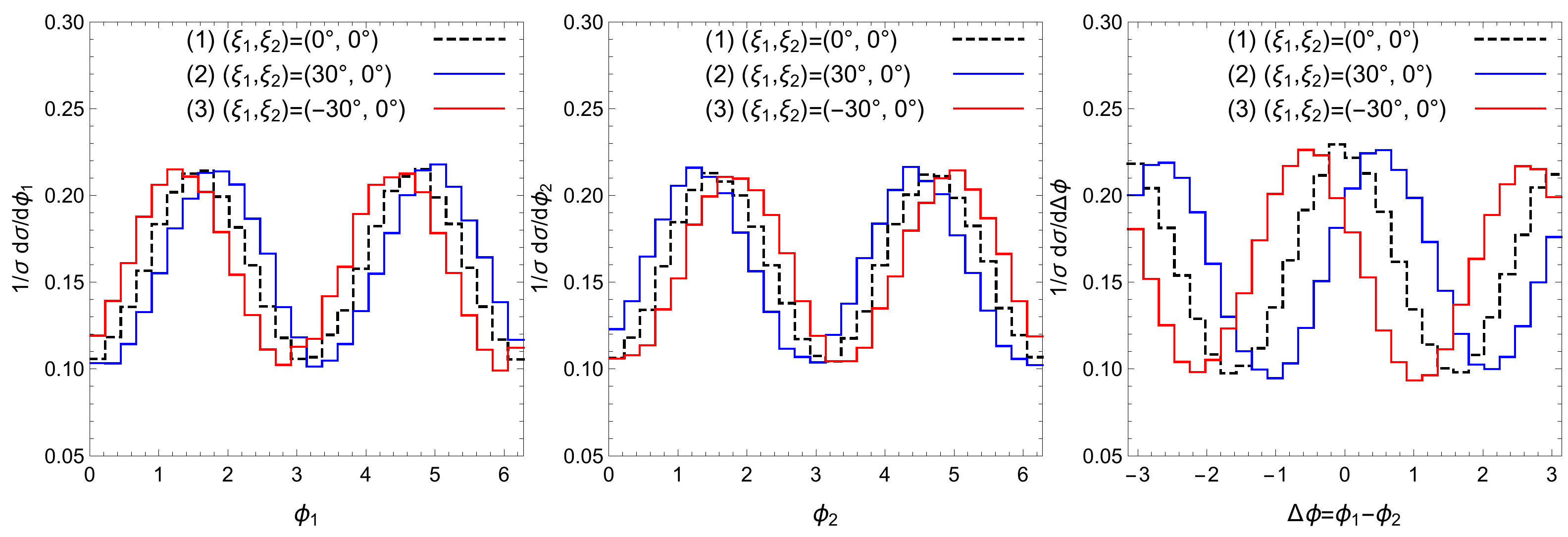}
\includegraphics[scale=0.5]{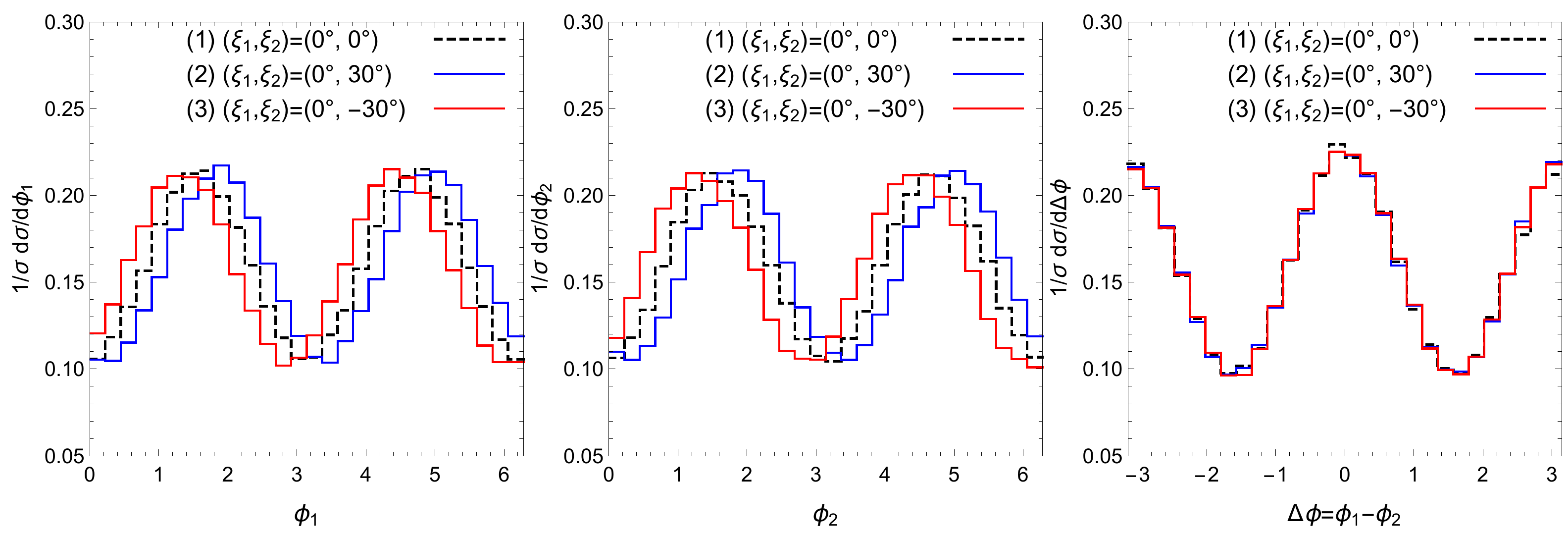}
\includegraphics[scale=0.5]{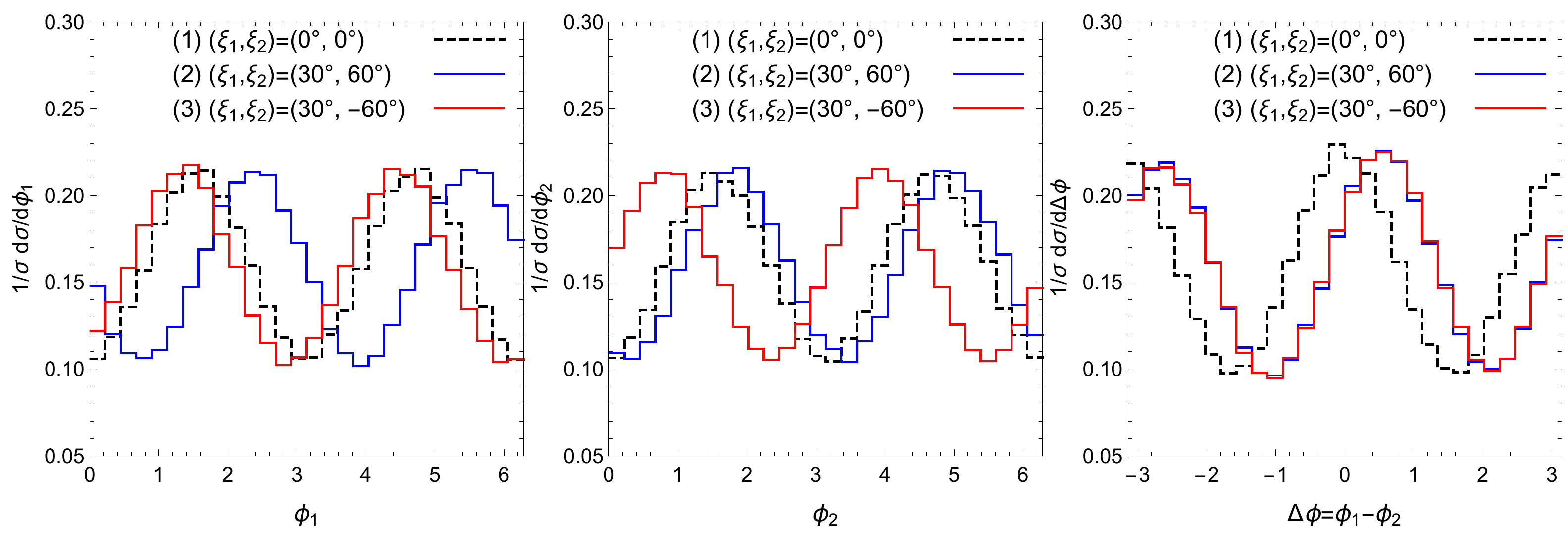}
\caption{\small 
The normalised differential cross section of the GF process as a function of $\phi_{1}^{}$ (left column), $\phi_{2}^{}$ (middle column) and $\Delta \phi$ (right column), with different values for the parity violating phases $\xi_{1,2}^{}$. In all of the panels, the SM prediction is shown by the black dashed curve. The correspondence between the curves and values for $\xi_{1,2}^{}$ is shown inside each panel. A cutoff on the $p_T^{}$ of the Higgs boson $p_T^{}>200$ GeV is imposed in the $\phi_{1}^{}$ and $\phi_{2}^{}$ panels. Each row has the same set of the parity violating phases.}
\label{figure:phi-dist-cpv}
\end{figure}

Finally, we study the impact of parity symmetry violation on the azimuthal angle correlations. Let us remind that we have introduced two phases $\xi_{1,2}^{}$ in eq.~(\ref{eq:cpv-phases}) and these phases parametrise the magnitude of parity violation in the process $gg\to HH$. If they are non-zero, the $gg\to HH$ amplitude $\hat{\cal M}_{\lambda_1^{} \lambda_2^{}}^{}$ is not parity invariant anymore: $\hat{\cal M}_{++}^{} \ne \hat{\cal M}_{--}^{}$ if $\xi_1^{}\ne0$, $\hat{\cal M}_{+-}^{} \ne \hat{\cal M}_{-+}^{}$ if $\xi_2^{}\ne0$. 
The squared VBF amplitude eq.~(\ref{full-amp-squared-ggh}) tells us that violation of the parity invariance of the amplitude appears as the following deviations from the standard model predictions: (1) the $\phi_1^{}$ and $\phi_2^{}$ distributions are not necessarily equal to each other, (2) the azimuthal angle observables do not necessarily show cosine distributions.
In order to make these expectations more explicit, we write the amplitude $\hat{\cal M}_{\lambda_1^{} \lambda_2^{}}^{}$ with the phases $\xi_{1,2}^{}$ in the following way:
\begin{align}
\hat{\cal M}_{\lambda, \lambda}^{} & = A\ e^{+i \lambda \xi_1^{}}_{}, \\
\hat{\cal M}_{\lambda, -\lambda}^{} & = B\ e^{+i \lambda \xi_2^{}}_{}.
\end{align}
These are simply obtained by putting eq.~(\ref{eq:cpv-phases}) into the amplitude in eq.~(\ref{amplitudes-gghh-2}) and substituting the terms including the form factors with $A$ and $B$. By inserting the above amplitudes into the azimuthal angle dependent terms in eq.~(\ref{full-amp-squared-ggh}), each term becomes
\begin{subequations}\label{eq:phase-dependence}
\begin{align}
-2 
\bigl[ &
{\it Re}\bigl( \hat{\cal M}_{++}^{} \hat{\cal M}_{-+}^* \bigr) +
{\it Re}\bigl( \hat{\cal M}_{+-}^{} \hat{\cal M}_{--}^* \bigr)
\bigr] \cos{2\phi_1^{}} + 
( {\it Re} \to {\it Im}, \cos{} \to \sin{} ) \nonumber \\
&=
-2\bigl( AB^*_{} + A^*_{}B \bigr) \cos{(2\phi_1^{} - \xi_1^{} - \xi_2^{})}, \\
\nonumber \\
-
2 
\bigl[ &
{\it Re}\bigl( \hat{\cal M}_{+-}^{} \hat{\cal M}_{++}^* \bigr) +
{\it Re}\bigl( \hat{\cal M}_{--}^{} \hat{\cal M}_{-+}^* \bigr)
\bigr] \cos{2\phi_2^{}} + 
( {\it Re} \to {\it Im}, \cos{} \to \sin{} ) \nonumber \\
&=
-2\bigl( AB^*_{} + A^*_{}B \bigr) \cos{(2\phi_2^{} + \xi_1^{} - \xi_2^{})}, \\
\nonumber \\
2 
{\it Re} & \bigl( \hat{\cal M}_{++}^{} \hat{\cal M}_{--}^* \bigr) 
\cos{2(\phi_1^{}-\phi_2^{})} +
( {\it Re} \to {\it Im}, \cos{} \to \sin{} ) 
= 2 |A|^2_{} \cos{2(\phi_1^{}-\phi_2^{}-\xi_1^{})}, \label{eq:phase-dependence-3}\\
2
{\it Re} & \bigl( \hat{\cal M}_{+-}^{} \hat{\cal M}_{-+}^* \bigr) 
\cos{2(\phi_1^{}+\phi_2^{})} +
( {\it Re} \to {\it Im}, \cos{} \to \sin{} ) 
= 2 |B|^2_{} \cos{2(\phi_1^{}+\phi_2^{}-\xi_2^{})}.
\end{align}
\end{subequations}
These results show that parity violation appears as peak shifts of the azimuthal angle distributions. 
The results also tell us an important fact that the peak shifts of the distributions reflect only the magnitude of parity violation (Recall that $\xi_{1,2}^{}$ parametrise the magnitude of parity violation in our study). 
It is an easy exercise to confirm that this is true no matter how parity violating phases are introduced. 
From the above analytic results, it is also apparent how each observable depends on the phases $\xi_{1,2}^{}$. $\Delta \phi$ is sensitive to $\xi_{1}^{}$ and $\phi_+^{}$ is sensitive to $\xi_{2}^{}$. $\phi_{1}^{}$ and $\phi_{2}^{}$ are sensitive to both of $\xi_{1}^{}$ and $\xi_{2}^{}$. Although we are far less likely to be able to measure $\xi_{2}^{}$ in the $\phi_+^{}$ distribution because of the very small correlation in $\phi_+^{}$, we may use $\phi_{1}^{}$ and $\phi_{2}^{}$ to probe $\xi_{2}^{}$ instead. \\

We note in passing the impact of parity violation in the process $pp\to Hjj$, which can be considered as a simpler case. 
Since the process $gg\to H$ has non-zero amplitudes $\hat{\cal M}_{\lambda_1^{} \lambda_2^{}}^{}$ only for $\Delta \lambda = \lambda_1^{}-\lambda_2^{} = 0$ states, the process $pp\to Hjj$ has only eq.~(\ref{eq:phase-dependence-3}) in its cross section formula. Therefore, parity violation in the process $gg\to H$ appears as a peak shift only in the $\Delta \phi$ distribution~\cite{Hankele:2006ma, Klamke:2007cu}.\\

While the phase dependence on the correlations is apparent from eq.~(\ref{eq:phase-dependence}), we present numerical results, too. 
In Figure~\ref{figure:phi-dist-cpv} we show the normalised differential cross section as a function of $\phi_{1}^{}$ (left column), $\phi_{2}^{}$ (middle column) and $\Delta \phi=\phi_1^{}-\phi_2^{}$ (right column) with different values for $\xi_{1,2}^{}$. The correspondence between the curves and values for $\xi_{1,2}^{}$ is shown inside each panel. 
In all of the panels, the SM prediction is shown by the black dashed curve. 
A cutoff, $p_T^{}>200$ GeV, is imposed on the $p_T^{}$ of the Higgs boson in the c.m. frame of the two Higgs bosons, when we produce the $\phi_{1}^{}$ and $\phi_{2}^{}$ plots, in order to enhance the correlations in $\phi_{1,2}^{}$, see Figure~\ref{figure:phi-dist-ptcut}. 
The distributions of $\phi_+^{}=\phi_1^{}+\phi_2^{}$ are not shown anymore, since we have found that the correlation in $\phi_+^{}$ is very small in most every case, see Figures~\ref{figure:phi-dist-ptcut} and \ref{figure:phi-dist-lambda}.

\subsection{The weak boson fusion process}\label{sec:wbf-process}

The weak boson fusion (WBF) sub-process $(V_1^{}, V_2^{})=(W^+_{}, W^-_{})$, $(W^-_{}, W^+_{})$ and $(Z, Z)$ consists of only the $qq$ initiated sub-process $(a_1^{}, a_2^{})=(q, q)$. 
However, the squared VBF amplitude for the WBF sub-process takes a more complicated form than that for the GF sub-process in eq.~(\ref{full-amp-squared-ggh}) because of the following two reasons: (1) the helicity $\lambda_{1,2}^{}=0$ components of the intermediate weak bosons additionally induce eight azimuthal angle dependent terms, such as $\cos{(\phi_1^{}-\phi_2^{})}$, (2) we cannot simply take averages for the initial helicities and take summations for the final helicities, since the electroweak interactions distinguish different helicity states. The squared VBF amplitude for four sets of the helicities of the external quarks must be prepared: $(\sigma_1^{}, \sigma_3^{}, \sigma_2^{}, \sigma_4^{})=(+,+,+,+), (+,+,-,-), (-,-,+,+), (-,-,-,-)$. \\

\begin{figure}
\centering
\includegraphics[scale=0.62]{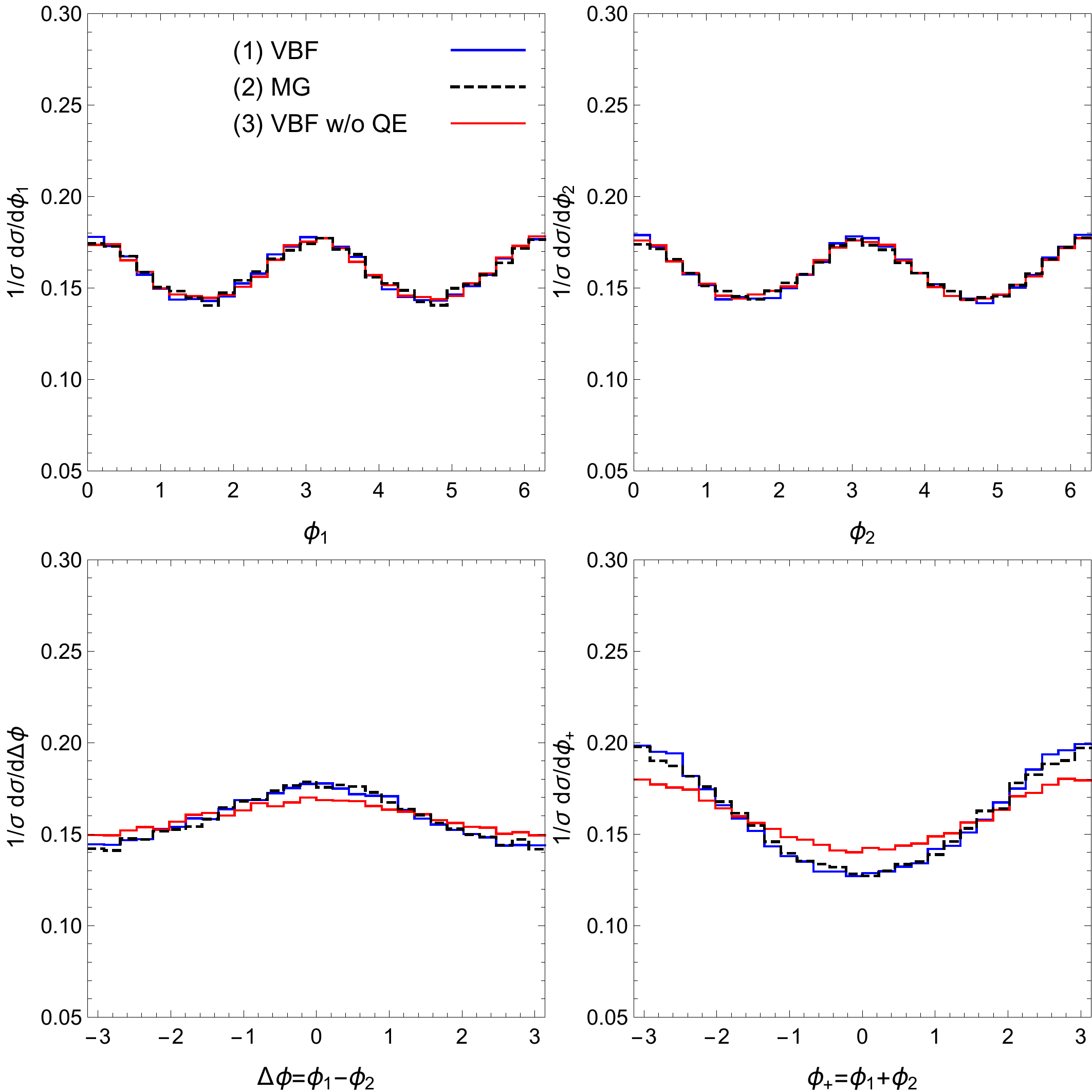}
\caption{\small 
The normalised differential cross section of the WBF process as a function of $\phi_1^{}$ (upper left), $\phi_2^{}$ (upper right), $\Delta \phi$ (lower left) and $\phi_+^{}$ (lower right). The correspondence between curves and simulation methods is shown inside the upper left panel: (1) The blue solid curve represents the result according to our analytic cross section formula, to which only the VBF diagram contributes, (2) The black dashed curve represents the exact LO result, (3) The red solid curve represents the result according to our analytic cross section formula, from which the quantum effects of the intermediate weak bosons are removed on purpose.}
\label{figure:phi-dist-weakfusion}
\end{figure}

The amplitude $( {\cal M}_{V_1^{}V_2^{}}^{HH} )_{\lambda_1^{} \lambda_2^{}}^{}$ for helicity $\lambda_{1,2}^{}=0$ weak bosons gives a dominant contribution in the WBF sub-process and the amplitudes for other helicities are much smaller than the above amplitude.
Thus we can expect that all of the azimuthal angle correlations are small, because the correlations arise from the interference of the amplitudes for various helicities, see eq.~(\ref{full-amp-squared-ggh}). 
If we keep only terms which contain at least one $( {\cal M}_{V_1^{}V_2^{}}^{HH} )_{0 0}^{}$, the squared VBF amplitude for $(\sigma_1^{}, \sigma_3^{}, \sigma_2^{}, \sigma_4^{})=(+,+,+,+)$ has the following form:
\begin{align}
\bigl| {\cal M}_{+ +}^{+ +} \bigr|^2_{}  = 
\frac{ 4 \bigl|g^{V_1^{}a_1^{}a_3^{}}_{+}\bigr|^2_{} \bigl|g^{V_2^{}a_2^{}a_4^{}}_{+}\bigr|^2_{}  }{ Q_1^2 Q_2^2 } 
& \Bigl\{
s_0^{} t_0^{} |\hat{\cal M}_{00}^{}|^2_{} \nonumber \\
& + 2 
\Bigl[
s_1^{} t_1^{} {\it Re}\bigl( \hat{\cal M}_{++}^{} \hat{\cal M}_{00}^* \bigr) +
s_2^{} t_2^{} {\it Re}\bigl( \hat{\cal M}_{00}^{} \hat{\cal M}_{--}^* \bigr)
\Bigr] \cos{(\phi_1^{}-\phi_2^{})} \nonumber \\
& - 2 
\Bigl[
s_1^{} t_2^{} {\it Re}\bigl( \hat{\cal M}_{+-}^{} \hat{\cal M}_{00}^* \bigr) +
s_2^{} t_1^{} {\it Re}\bigl( \hat{\cal M}_{00}^{} \hat{\cal M}_{-+}^* \bigr)
\Bigr] \cos{(\phi_1^{}+\phi_2^{})} \nonumber \\
& + 
\bigl( {\it Re} \to {\it Im}, \cos{} \to \sin{} \bigr)
\Bigr\},\label{full-amp-squared-wwh}
\end{align}
where we introduce a notation $\hat{\cal M}_{\lambda_1^{} \lambda_2^{}}^{}$ which denotes the amplitude $( {\cal M}_{V_1^{}V_2^{}}^{HH} )_{\lambda_1^{} \lambda_2^{}}^{}$. $s_{0,1,2}^{}$ and $t_{0,1,2}^{}$ are functions of $\theta_{1,2}^{}$ defined in the $q_{1,2}^{}$ Breit frames eqs.~(\ref{eq:breit1}) and (\ref{eq:breit2}) and given by
\begin{subequations}
\begin{align}
s_0^{} = \frac{ \sin^2_{}{\theta_1^{}} }{2 \cos^2_{}{\theta_1^{}} },\ \ 
s_1^{} = \frac{ \sqrt{2} \sin{\theta_1^{}} (1+\cos{\theta_1^{}}) }{4 \cos^2_{}{\theta_1^{}} },\ \ 
s_2^{} = \frac{ \sqrt{2} \sin{\theta_1^{}} (1-\cos{\theta_1^{}}) }{4 \cos^2_{}{\theta_1^{}} }, \\
t_0^{} = \frac{ \sin^2_{}{\theta_2^{}} }{2 \cos^2_{}{\theta_2^{}} },\ \ 
t_1^{} = \frac{ \sqrt{2} \sin{\theta_2^{}} (1-\cos{\theta_2^{}}) }{4 \cos^2_{}{\theta_2^{}} },\ \ 
t_2^{} = \frac{ \sqrt{2} \sin{\theta_2^{}} (1+\cos{\theta_2^{}}) }{4 \cos^2_{}{\theta_2^{}} }.
\end{align}
\end{subequations}
The squared VBF amplitude for the other three helicity states can be simply obtained by exchanging $s_{1,2}^{}$ and $t_{1,2}^{}$ in $| {\cal M}_{+ +}^{+ +} |^2_{}$ in the following ways: 
\begin{align}
| {\cal M}_{+ +}^{+ +} |^2_{} \to | {\cal M}_{- +}^{- +} |^2_{} \ \ & \text{by}~ s_1^{} \leftrightarrow s_2^{},\,\nonumber\\
| {\cal M}_{+ +}^{+ +} |^2_{} \to | {\cal M}_{+ -}^{+ -} |^2_{} \ \ & \text{by}~ t_1^{} \leftrightarrow t_2^{},\,\nonumber\\
| {\cal M}_{+ +}^{+ +} |^2_{} \to | {\cal M}_{- -}^{- -} |^2_{}  \ \ & \text{by}~ s_1^{} \leftrightarrow s_2^{}~\text{and}~t_1^{} \leftrightarrow t_2^{}.
\end{align}
The couplings should be also changed accordingly. 
The coefficients of $\cos{\phi_{1,2}^{}}$ terms actually contain one $\hat{\cal M}_{00}^{}$, too. However, the $\cos{\phi_{1,2}^{}}$ terms cannot give correlated distributions in the process $pp\to HHjj$, because we cannot distinguish the two Higgs bosons. 
More practically, an azimuthal angle dependent term which changes its overall sign under the transformations $\phi_1^{} \to \phi_1^{}+\pi$ and $\phi_2^{} \to \phi_2^{}+\pi$ gives only a flat distribution after the phase space integration. 
The first term in the right hand side (RHS) of eq.~(\ref{full-amp-squared-wwh}) contributes to the inclusive cross section after the phase space integration and the other terms give the correlations in $\Delta \phi=\phi_1^{}-\phi_2^{}$ and $\phi_+^{}=\phi_1^{}-\phi_2^{}$. 
An interesting difference from the GF sub-process is that the sine terms do not vanish even when the amplitude is parity invariant ($\hat{\cal M}_{++}^{} = \hat{\cal M}_{--}^{}$ and $\hat{\cal M}_{+-}^{} = \hat{\cal M}_{-+}^{}$), if the amplitude contains an imaginary part. This is because the interaction between the external quarks and the intermediate weak boson already violates the parity symmetry. In our tree-level calculation, the amplitude is purely real and we will observe only cosine distributions. \\

We show numerical results for the 14 TeV LHC. The setup and phase space cuts are the same as in the GF study in Section~\ref{sec:gf-process}. Therefore the numerical results in this Section can be directly compared with those in Section~\ref{sec:gf-process}. The only difference is the scale choice in the PDFs. The $p_T^{}$ of the jet with a positive rapidity is used for the scale in the PDF of the incoming parton which moves along the positive direction of the $z$-axis, and the $p_T^{}$ of the jet with a negative rapidity is used for the scale in the PDF of the other incoming parton. 
In Figure~\ref{figure:phi-dist-weakfusion}, we show the normalised differential cross section as a function of $\phi_1^{}$ (upper left), $\phi_2^{}$ (upper right), $\Delta \phi = \phi_1^{} - \phi_2^{}$ (lower left) and $\phi_+^{} = \phi_1^{} + \phi_2^{}$ (lower right). The blue solid curve, labelled as (1) VBF, represents the result according to our analytic cross section formula, to which only the VBF diagrams contribute. The black dashed curve, labelled as (2) MG, represents the result according to the exact LO cross section, to which not only the VBF diagrams but also the $s$-channel and $u$-channel diagrams contribute, generated by {\tt MadGraph5\verb|_|aMC@NLO}\cite{Alwall:2014hca} version 5.2.2.1. 
The red solid curve, labelled as (3) VBF w/o QE, represents the result according to our analytic cross section formula from which the azimuthal angle dependent terms are removed on purpose, that is, the quantum effects of the intermediate weak bosons expressed as the interference of the amplitudes for different helicities are killed. Therefore, the differences between the result (1) and the result (3) in the distributions visualise the contribution from the azimuthal angle dependent terms. 
The good agreement between the result (1) and the result (2) confirms the validity of our analytic cross section formula. 
The $\phi_1^{}$ and $\phi_2^{}$ plots show correlated distributions. However, the agreement between the result (1) and the result (3) indicates that the correlated distributions are not induced by the quantum effects of the intermediate weak bosons but by a kinematic effect. 
The small discrepancies between the result (1) and the result (3) in the $\Delta \phi$ and $\phi_+^{}$ plots come from the $\cos{\Delta \phi}$ and the $\cos{\phi_+^{}}$ terms in eq.~(\ref{full-amp-squared-wwh}), respectively. The smallness of the discrepancies is as expected (see the discussion above eq.~(\ref{full-amp-squared-wwh})). In the $\Delta \phi$ and $\phi_+^{}$ plots, the result (3) again shows correlated distributions, which must be induced by a kinematic effect. 
Therefore, the WBF sub-process produces the correlated distributions in all of the azimuthal angle observables, however they are mainly induced by a kinematic effect.
We note that the GF sub-process produces only flat distributions in all of the azimuthal angle observables, when the quantum effects of the intermediate gluons are killed in the same way as above. It can be concluded that the non-flat distributions induced by a kinematic effect are a characteristic feature of the WBF sub-process. \\

\begin{figure}[t]
\centering
\includegraphics[scale=0.5]{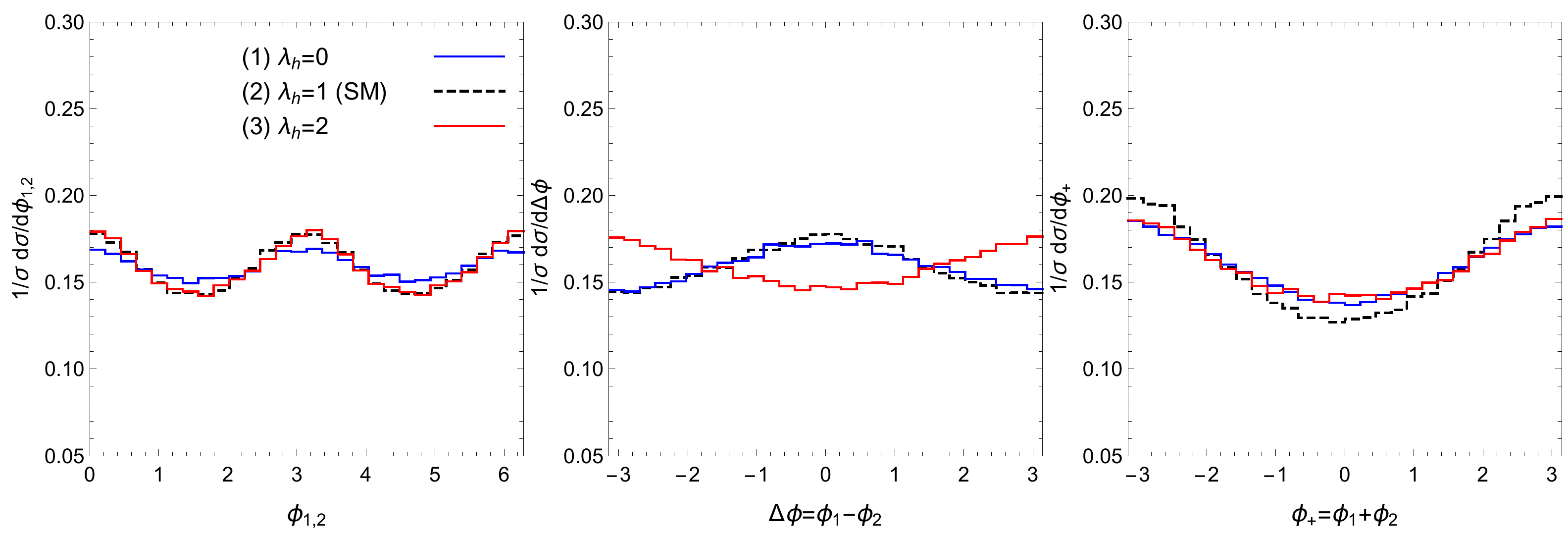}
\caption{\small 
The normalised differential cross section of the WBF process as a function of $\phi_{1,2}^{}$ (left), $\Delta \phi$ (middle) and $\phi_+^{}$ (right), with three different values for the triple Higgs self-coupling re-scaling factor $\lambda_h^{}$. The correspondence between the curves and values for $\lambda_h^{}$ is shown inside the left panel.}
\label{figure:phi-dist-weakfusion-lam}
\end{figure}

We study how the azimuthal angle distributions depend on the triple Higgs self-coupling. We have observed the kinematic effect on the distributions in non-standard cases $\lambda_h^{} \ne 1$, too, where $\lambda_h^{}$ is the factor re-scaling the triple Higgs self-coupling. 
The three cases $\lambda_h^{}=0, 1, 2$ produce the similar distributions in the azimuthal angle observables, when the quantum effects of the intermediate weak bosons are killed.
In Figure~\ref{figure:phi-dist-weakfusion-lam} we show the normalised differential cross section as a function of $\phi_{1,2}^{}$ (left), $\Delta \phi = \phi_1^{} - \phi_2^{}$ (middle) and $\phi_+^{} = \phi_1^{} + \phi_2^{}$ (right) with three different values of $\lambda_h^{}$, the blue solid curve: $\lambda_h^{}=0$, the black dashed curve: $\lambda_h^{}=1$ (the SM prediction), and the red solid curve: $\lambda_h^{}=2$. 
The correlated distributions in the $\phi_{1,2}^{}$ plot are completely induced by a kinematic effect. 
We find that, when $\lambda_h^{}=2$, the coefficient of the $\cos{(\phi_1^{}-\phi_2^{})}$ term (the second term in the RHS of eq.~(\ref{full-amp-squared-wwh})) is large enough to flip the $\Delta \phi$ distribution. The impact of a non-standard value for $\lambda_h^{}$ ($\lambda_h^{} \ne 1$) is again not so significant. However, differently from the GF sub-process which is actually the $\mathcal{O}(\alpha_s^2)$ correction to the inclusive GF sub-process $gg\to HH$, the WBF sub-process is a LO tree-level process and so the correlations should be used together with other observables, such as the invariant mass of the Higgs boson pair, to probe $\lambda_h^{}$.

\section{Summary and discussion}\label{sec:conclusion}

In this paper, we have studied the azimuthal angle correlations of two jets in the production of a Higgs boson pair plus two jets $pp\to HHjj$. Based on the known fact that the azimuthal angle correlations are induced by the quantum effects of the two intermediate vector bosons in vector boson fusion (VBF) sub-processes, we have calculated the amplitudes contributed from only VBF Feynman diagrams. As VBF sub-processes, we have considered the gluon fusion (GF) sub-process, which is an one-loop $\mathcal{O}(\alpha_s^4 \alpha_{}^2)$ process at leading order (LO), and the weak boson fusion (WBF) sub-process, which is an $\mathcal{O}(\alpha_{}^4)$ process at LO. We have used a helicity amplitude technique for evaluating the VBF amplitudes. Based on the method presented in ref.~\cite{Hagiwara:2009wt}, we have divided a VBF amplitude into two amplitudes for off-shell vector boson emissions ($q\to qV^*_{}$ or $g\to gV^*_{}$) and one amplitude for the Higgs boson pair production by a fusion of the two off-shell vector bosons ($V_{}^*V_{}^* \to HH$), and presentd each of the three amplitudes in the helicity basis. 
The quantum effects of the intermediate vector bosons are still included correctly and are expressed as the interference of the $V_{}^*V_{}^* \to HH$ amplitudes with various helicities of the vector bosons. With this method, we have obtained the analytic cross section formula in a compact form, from which we can easily make an expectation on the azimuthal angle correlations, both for the GF sub-process and for the WBF sub-process. We have numerically compared our analytic cross section formulas with the exact LO results and have observed the good agreement between the two results both in the inclusive cross sections and in azimuthal angle distributions, after the VBF cuts and the upper transverse momentum $p_T^{}$ cut on the jets are imposed (For the GF sub-process, the comparison is performed only in the large $m_t^{}$ limit.). As azimuthal angle observables, we have studied four observables: $\phi_1^{}$, $\phi_2^{}$, $\Delta \phi = \phi_1^{} - \phi_2^{}$ and $\phi_+^{} = \phi_1^{} + \phi_2^{}$, where $\phi_{1,2}^{}$ are the azimuthal angles of the two jets measured from the production plane of the Higgs boson pair.\\

In the GF sub-process, using a finite $m_t^{}$ value is found to be important to produce the azimuthal angle correlations correctly. 
The GF sub-process exhibits large correlations in $\phi_{1,2}^{}$ and $\Delta \phi$. The $p_T^{}$ of the Higgs boson is found to be useful in controlling these correlations. The correlation in $\Delta \phi$ is enhanced when the $p_T^{}$ of the Higgs boson is decreased and the correlations in $\phi_{1,2}^{}$ are enhanced when the $p_T^{}$ of the Higgs boson is increased. 
We have found that the correlation in $\phi_+^{}$ is very small in most every case.
The impact of a non-standard value for the triple Higgs self-coupling on the correlations is found to be much smaller than that in other observables, such as the invariant mass of the Higgs boson pair, of the inclusive process $pp\to HH$. 
In order to study the impact of parity violation on the correlations, 
we have introduced two independent phases $\xi_{1,2}^{}$ which parametrise the magnitude of parity violation in the process $gg \to HH$, in a way that $\xi_{1}^{}$ affects the $gg \to HH$ amplitude for $\Delta \lambda=0$ helicity states, where $\Delta \lambda=\lambda_1^{}-\lambda_2^{}$ and $\lambda_{1,2}^{}$ are helicities of the gluons, and $\xi_{2}^{}$ affects the amplitude for $\Delta \lambda=\pm 2$ helicity states.
We have analytically shown that parity violation appears as peak shifts of the correlations and that the peak shifts reflect only the magnitude of parity violation. We have also shown that $\Delta \phi$ is sensitive to $\xi_1^{}$, $\phi_+^{}$ is sensitive to $\xi_2^{}$, and $\phi_{1,2}^{}$ are sensitive to both of $\xi_1^{}$ and $\xi_2^{}$. Although we are far less likely to be able to measure $\xi_{2}^{}$ in the $\phi_+^{}$ distribution because of the very small correlation in $\phi_+^{}$, we may use $\phi_{1}^{}$ and $\phi_{2}^{}$ to probe $\xi_{2}^{}$ instead. 
While we can naively expect that the azimuthal angle distributions in the WBF sub-process are almost flat, we have actually observed correlated (non-flat) distributions. 
We have found that they are not induced by the quantum effects of the intermediate weak bosons but by a kinematic effect. Since we do not find the similar kinematic effect in the GF sub-process, we conclude that this is a characteristic feature of the WBF sub-process. The impact of a non-standard value for the triple Higgs self-coupling is not so significant in the WBF sub-process, too. \\

The parton level event samples of the process $pp\to HHjj$ are exclusively generated and each of the two outgoing partons is identified as a jet in our numerical studies. When more realistic event generations are intended to be performed, merging the parton level event samples with the leading logarithmic parton shower~\cite{Catani:2001cc, Lonnblad:2001iq, Alwall:2007fs} and subsequently proceeding to a hadronisation procedure will be a promising approach. However, a careful merging procedure is required for correctly reproducing the azimuthal angle correlations after the merging procedure, because the correlations studied in this paper are completely process dependent ($pp\to HHjj$) and those process dependent angular correlations are not described correctly by the parton shower. Contamination from the parton shower may lead to a wrong prediction~\cite{Nakamura:2015uca}. \\

\begin{figure}[t]
\centering
\includegraphics[scale=0.5]{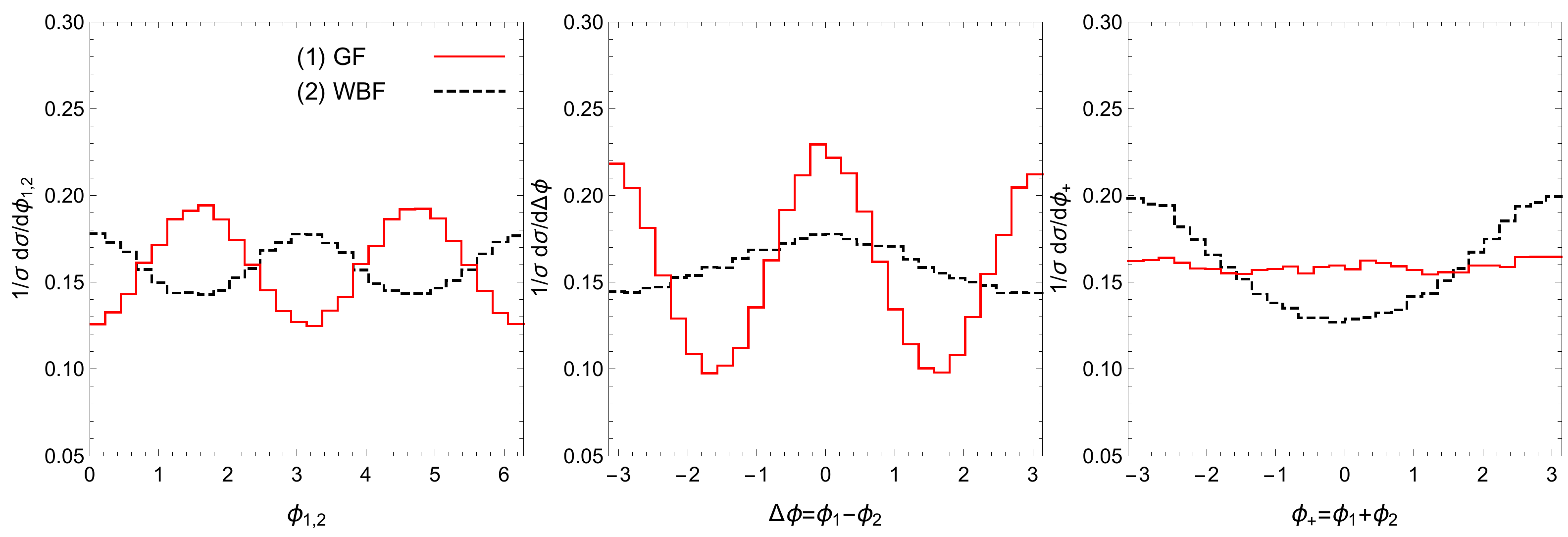}
\caption{\small 
The normalised differential cross section as a function of $\phi_{1,2}^{}$ (left), $\Delta \phi$ (middle) and $\phi_+^{}$ (right) for the GF process (red solid curve) and the WBF process (black dashed curve).}
\label{figure:phi-dist-gf-wbf}
\end{figure}

The azimuthal angle correlations revealed in this paper will help the analyses of the process $pp\to HHjj$ at the LHC. 
Since the production cross section of the process $pp\to HHjj$ is small, we should use as much process dependent information as possible to extract the events of the process. 
The azimuthal angle correlations are obviously a part of the process dependent information. 
It is a known issue that separating the contributions coming from the GF sub-process and those coming from the WBF sub-process is difficult in the production of a Higgs boson pair plus two jets~\cite{Dolan:2013rja}. 
One possible application of the correlations is to help disentangling the GF sub-process and the WBF sub-process by using the fact that these two sub-processes exhibit different correlations, as shown in Figure~\ref{figure:phi-dist-gf-wbf}.
We have studied only the signal processes in this paper and so the impact of the correlations in a realistic situation is not so clear yet. 
The fully automated event generation for loop induced processes is now available in MadGraph5\verb|_|aMC@NLO~\cite{Alwall:2014hca, Hirschi:2015iia}. This achievement will activate phenomenological studies of the process. We hope that further phenomenological studies including the uses of the azimuthal angle correlations will be performed both by theorists and by experimentalists.

\section*{Acknowledgments}

J.N. is grateful to Kentarou Mawatari for answering several questions on physics of jet angular correlations. J.N. would also like to thank Kaoru Hagiwara for valuable discussions, and Junichi Kanzaki for his help on using pragrams {\tt BASES} and {\tt SPRING}. The authors would also like to thank Barbara J\"ager for useful discussions. 
The work of the authors (J.N. and J.B.) is supported by the Institutional Strategy of the University of T\"ubingen (DFG, ZUK 63) and in addition J.B. is supported by the DFG Grant JA 1954/1.


\bibliographystyle{JHEP}
\nocite{*}
\bibliography{HHjj_bib}

\providecommand{\href}[2]{#2}\begingroup\raggedright\begin{thebibliography}{10}

\bibitem{Aad:2012tfa}
{\bf ATLAS} Collaboration, G.~Aad {\em et.~al.}, {\it {Observation of a new
  particle in the search for the Standard Model Higgs boson with the ATLAS
  detector at the LHC}},  {\em Phys. Lett.} {\bf B716} (2012) 1--29,
  [\href{http://xxx.lanl.gov/abs/1207.7214}{{\tt arXiv:1207.7214}}].

\bibitem{Chatrchyan:2012xdj}
{\bf CMS} Collaboration, S.~Chatrchyan {\em et.~al.}, {\it {Observation of a
  new boson at a mass of 125 GeV with the CMS experiment at the LHC}},  {\em
  Phys. Lett.} {\bf B716} (2012) 30--61,
  [\href{http://xxx.lanl.gov/abs/1207.7235}{{\tt arXiv:1207.7235}}].

\bibitem{Aad:2014eva}
{\bf ATLAS} Collaboration, G.~Aad {\em et.~al.}, {\it {Measurements of Higgs
  boson production and couplings in the four-lepton channel in pp collisions at
  center-of-mass energies of 7 and 8 TeV with the ATLAS detector}},  {\em
  Phys.Rev.} {\bf D91} (2015), no.~1 012006,
  [\href{http://xxx.lanl.gov/abs/1408.5191}{{\tt arXiv:1408.5191}}].

\bibitem{Khachatryan:2014kca}
{\bf CMS} Collaboration, V.~Khachatryan {\em et.~al.}, {\it {Constraints on the
  spin-parity and anomalous HVV couplings of the Higgs boson in proton
  collisions at 7 and 8 TeV}},  \href{http://xxx.lanl.gov/abs/1411.3441}{{\tt
  arXiv:1411.3441}}.

\bibitem{ATLAS-CONF-2015-008}
{\bf ATLAS} Collaboration, G.~Aad {\em et.~al.}, {\it {Study of the spin and
  parity of the Higgs boson in HVV decays with the ATLAS detector}},  Tech.
  Rep. ATLAS-CONF-2015-008, CERN, 2015.

\bibitem{Higgs:1964ia}
P.~W. Higgs, {\it {Broken symmetries, massless particles and gauge fields}},
  {\em Phys.Lett.} {\bf 12} (1964) 132--133.

\bibitem{Englert:1964et}
F.~Englert and R.~Brout, {\it {Broken Symmetry and the Mass of Gauge Vector
  Mesons}},  {\em Phys.Rev.Lett.} {\bf 13} (1964) 321--323.

\bibitem{Higgs:1964pj}
P.~W. Higgs, {\it {Broken Symmetries and the Masses of Gauge Bosons}},  {\em
  Phys.Rev.Lett.} {\bf 13} (1964) 508--509.

\bibitem{Guralnik:1964eu}
G.~Guralnik, C.~Hagen, and T.~Kibble, {\it {Global Conservation Laws and
  Massless Particles}},  {\em Phys.Rev.Lett.} {\bf 13} (1964) 585--587.

\bibitem{Arkani-Hamed:2015vfh}
N.~Arkani-Hamed, T.~Han, M.~Mangano, and L.-T. Wang, {\it {Physics
  Opportunities of a 100 TeV Proton-Proton Collider}},
  \href{http://xxx.lanl.gov/abs/1511.0649}{{\tt arXiv:1511.0649}}.

\bibitem{Baglio:2015wcg}
J.~Baglio, A.~Djouadi, and J.~Quevillon, {\it {Prospects for Higgs physics at
  energies up to 100 TeV}},  \href{http://xxx.lanl.gov/abs/1511.0785}{{\tt
  arXiv:1511.0785}}.

\bibitem{Djouadi:1999gv}
A.~Djouadi, W.~Kilian, M.~Muhlleitner, and P.~M. Zerwas, {\it {Testing Higgs
  selfcouplings at e+ e- linear colliders}},  {\em Eur. Phys. J.} {\bf C10}
  (1999) 27--43, [\href{http://xxx.lanl.gov/abs/hep-ph/9903229}{{\tt
  hep-ph/9903229}}].

\bibitem{Djouadi:1999rca}
A.~Djouadi, W.~Kilian, M.~Muhlleitner, and P.~M. Zerwas, {\it {Production of
  neutral Higgs boson pairs at LHC}},  {\em Eur. Phys. J.} {\bf C10} (1999)
  45--49, [\href{http://xxx.lanl.gov/abs/hep-ph/9904287}{{\tt
  hep-ph/9904287}}].

\bibitem{Baur:2002rb}
U.~Baur, T.~Plehn, and D.~L. Rainwater, {\it {Measuring the Higgs boson self
  coupling at the LHC and finite top mass matrix elements}},  {\em Phys. Rev.
  Lett.} {\bf 89} (2002) 151801,
  [\href{http://xxx.lanl.gov/abs/hep-ph/0206024}{{\tt hep-ph/0206024}}].

\bibitem{Baur:2002qd}
U.~Baur, T.~Plehn, and D.~L. Rainwater, {\it {Determining the Higgs boson
  selfcoupling at hadron colliders}},  {\em Phys. Rev.} {\bf D67} (2003)
  033003, [\href{http://xxx.lanl.gov/abs/hep-ph/0211224}{{\tt
  hep-ph/0211224}}].

\bibitem{Baur:2003gpa}
U.~Baur, T.~Plehn, and D.~L. Rainwater, {\it {Examining the Higgs boson
  potential at lepton and hadron colliders: A Comparative analysis}},  {\em
  Phys. Rev.} {\bf D68} (2003) 033001,
  [\href{http://xxx.lanl.gov/abs/hep-ph/0304015}{{\tt hep-ph/0304015}}].

\bibitem{Baur:2003gp}
U.~Baur, T.~Plehn, and D.~L. Rainwater, {\it {Probing the Higgs selfcoupling at
  hadron colliders using rare decays}},  {\em Phys. Rev.} {\bf D69} (2004)
  053004, [\href{http://xxx.lanl.gov/abs/hep-ph/0310056}{{\tt
  hep-ph/0310056}}].

\bibitem{Dolan:2012rv}
M.~J. Dolan, C.~Englert, and M.~Spannowsky, {\it {Higgs self-coupling
  measurements at the LHC}},  {\em JHEP} {\bf 10} (2012) 112,
  [\href{http://xxx.lanl.gov/abs/1206.5001}{{\tt arXiv:1206.5001}}].

\bibitem{Baglio:2012np}
J.~Baglio, A.~Djouadi, R.~{Gr\"{o}ber}, M.~M. {M\"{u}hlleitner}, J.~Quevillon,
  and M.~Spira, {\it {The measurement of the Higgs self-coupling at the LHC:
  theoretical status}},  {\em JHEP} {\bf 04} (2013) 151,
  [\href{http://xxx.lanl.gov/abs/1212.5581}{{\tt arXiv:1212.5581}}].

\bibitem{Papaefstathiou:2012qe}
A.~Papaefstathiou, L.~L. Yang, and J.~Zurita, {\it {Higgs boson pair production
  at the LHC in the $b \bar{b} W^+ W^-$ channel}},  {\em Phys. Rev.} {\bf D87}
  (2013), no.~1 011301, [\href{http://xxx.lanl.gov/abs/1209.1489}{{\tt
  arXiv:1209.1489}}].

\bibitem{Barr:2013tda}
A.~J. Barr, M.~J. Dolan, C.~Englert, and M.~Spannowsky, {\it {Di-Higgs final
  states augMT2ed -- selecting $hh$ events at the high luminosity LHC}},  {\em
  Phys. Lett.} {\bf B728} (2014) 308--313,
  [\href{http://xxx.lanl.gov/abs/1309.6318}{{\tt arXiv:1309.6318}}].

\bibitem{Barger:2013jfa}
V.~Barger, L.~L. Everett, C.~B. Jackson, and G.~Shaughnessy, {\it {Higgs-Pair
  Production and Measurement of the Triscalar Coupling at LHC(8,14)}},  {\em
  Phys. Lett.} {\bf B728} (2014) 433--436,
  [\href{http://xxx.lanl.gov/abs/1311.2931}{{\tt arXiv:1311.2931}}].

\bibitem{Dolan:2013rja}
M.~J. Dolan, C.~Englert, N.~Greiner, and M.~Spannowsky, {\it {Further on up the
  road: $hhjj$ production at the LHC}},  {\em Phys. Rev. Lett.} {\bf 112}
  (2014) 101802, [\href{http://xxx.lanl.gov/abs/1310.1084}{{\tt
  arXiv:1310.1084}}].

\bibitem{deFlorian:2013uza}
D.~de~Florian and J.~Mazzitelli, {\it {Two-loop virtual corrections to Higgs
  pair production}},  {\em Phys. Lett.} {\bf B724} (2013) 306--309,
  [\href{http://xxx.lanl.gov/abs/1305.5206}{{\tt arXiv:1305.5206}}].

\bibitem{deFlorian:2013jea}
D.~de~Florian and J.~Mazzitelli, {\it {Higgs Boson Pair Production at
  Next-to-Next-to-Leading Order in QCD}},  {\em Phys. Rev. Lett.} {\bf 111}
  (2013) 201801, [\href{http://xxx.lanl.gov/abs/1309.6594}{{\tt
  arXiv:1309.6594}}].

\bibitem{Goertz:2014qta}
F.~Goertz, A.~Papaefstathiou, L.~L. Yang, and J.~Zurita, {\it {Higgs boson pair
  production in the D=6 extension of the SM}},  {\em JHEP} {\bf 04} (2015) 167,
  [\href{http://xxx.lanl.gov/abs/1410.3471}{{\tt arXiv:1410.3471}}].

\bibitem{deLima:2014dta}
D.~E. Ferreira~de Lima, A.~Papaefstathiou, and M.~Spannowsky, {\it {Standard
  model Higgs boson pair production in the ( $ b\overline{b} $ )( $
  b\overline{b} $ ) final state}},  {\em JHEP} {\bf 08} (2014) 030,
  [\href{http://xxx.lanl.gov/abs/1404.7139}{{\tt arXiv:1404.7139}}].

\bibitem{Frederix:2014hta}
R.~Frederix, S.~Frixione, V.~Hirschi, F.~Maltoni, O.~Mattelaer, P.~Torrielli,
  E.~Vryonidou, and M.~Zaro, {\it {Higgs pair production at the LHC with NLO
  and parton-shower effects}},  {\em Phys. Lett.} {\bf B732} (2014) 142--149,
  [\href{http://xxx.lanl.gov/abs/1401.7340}{{\tt arXiv:1401.7340}}].

\bibitem{Maltoni:2014eza}
F.~Maltoni, E.~Vryonidou, and M.~Zaro, {\it {Top-quark mass effects in double
  and triple Higgs production in gluon-gluon fusion at NLO}},  {\em JHEP} {\bf
  11} (2014) 079, [\href{http://xxx.lanl.gov/abs/1408.6542}{{\tt
  arXiv:1408.6542}}].

\bibitem{Azatov:2015oxa}
A.~Azatov, R.~Contino, G.~Panico, and M.~Son, {\it {Effective field theory
  analysis of double Higgs boson production via gluon fusion}},  {\em Phys.
  Rev.} {\bf D92} (2015), no.~3 035001,
  [\href{http://xxx.lanl.gov/abs/1502.0053}{{\tt arXiv:1502.0053}}].

\bibitem{Dolan:2015zja}
M.~J. Dolan, C.~Englert, N.~Greiner, K.~Nordstrom, and M.~Spannowsky, {\it
  {$hhjj$ production at the LHC}},  {\em Eur. Phys. J.} {\bf C75} (2015), no.~8
  387, [\href{http://xxx.lanl.gov/abs/1506.0800}{{\tt arXiv:1506.0800}}].

\bibitem{deFlorian:2015moa}
D.~de~Florian and J.~Mazzitelli, {\it {Higgs pair production at
  next-to-next-to-leading logarithmic accuracy at the LHC}},  {\em JHEP} {\bf
  09} (2015) 053, [\href{http://xxx.lanl.gov/abs/1505.0712}{{\tt
  arXiv:1505.0712}}].

\bibitem{Barr:2014sga}
A.~J. Barr, M.~J. Dolan, C.~Englert, D.~E. Ferreira~de Lima, and M.~Spannowsky,
  {\it {Higgs Self-Coupling Measurements at a 100 TeV Hadron Collider}},  {\em
  JHEP} {\bf 02} (2015) 016, [\href{http://xxx.lanl.gov/abs/1412.7154}{{\tt
  arXiv:1412.7154}}].

\bibitem{Papaefstathiou:2015iba}
A.~Papaefstathiou, {\it {Discovering Higgs boson pair production through rare
  final states at a 100 TeV collider}},  {\em Phys. Rev.} {\bf D91} (2015),
  no.~11 113016, [\href{http://xxx.lanl.gov/abs/1504.0462}{{\tt
  arXiv:1504.0462}}].

\bibitem{Kotwal:2015rba}
A.~V. Kotwal, S.~Chekanov, and M.~Low, {\it {Double Higgs Boson Production in
  the 4$\tau$ Channel from Resonances in Longitudinal Vector Boson Scattering
  at a 100 TeV Collider}},  {\em Phys. Rev.} {\bf D91} (2015) 114018,
  [\href{http://xxx.lanl.gov/abs/1504.0804}{{\tt arXiv:1504.0804}}].

\bibitem{Eboli:1987dy}
O.~J.~P. {\'Eboli}, G.~C. Marques, S.~F. Novaes, and A.~A. Natale, {\it {Twin
  Higgs Boson Production}},  {\em Phys. Lett.} {\bf B197} (1987) 269.

\bibitem{Glover:1987nx}
E.~W.~N. Glover and J.~J. van~der Bij, {\it {Higgs Boson Pair Production Via
  Gluon Fusion}},  {\em Nucl. Phys.} {\bf B309} (1988) 282.

\bibitem{Dicus:1987ic}
D.~A. Dicus, C.~Kao, and S.~S.~D. Willenbrock, {\it {Higgs Boson Pair
  Production From Gluon Fusion}},  {\em Phys. Lett.} {\bf B203} (1988) 457.

\bibitem{Plehn:1996wb}
T.~Plehn, M.~Spira, and P.~M. Zerwas, {\it {Pair production of neutral Higgs
  particles in gluon-gluon collisions}},  {\em Nucl. Phys.} {\bf B479} (1996)
  46--64, [\href{http://xxx.lanl.gov/abs/hep-ph/9603205}{{\tt
  hep-ph/9603205}}]. [Erratum: Nucl. Phys.B531,655(1998)].

\bibitem{Keung:1987nw}
W.-Y. Keung, {\it {Double Higgs From $W - W$ Fusion}},  {\em Mod. Phys. Lett.}
  {\bf A2} (1987) 765.

\bibitem{Dicus:1987ez}
D.~A. Dicus, K.~J. Kallianpur, and S.~S.~D. Willenbrock, {\it {Higgs Boson Pair
  Production in the Effective $W$ Approximation}},  {\em Phys. Lett.} {\bf
  B200} (1988) 187.

\bibitem{Dobrovolskaya:1990kx}
A.~Dobrovolskaya and V.~Novikov, {\it {On heavy Higgs boson production}},  {\em
  Z. Phys.} {\bf C52} (1991) 427--436.

\bibitem{Baglio:2014aka}
J.~Baglio, {\it {A theoretical review of triple Higgs coupling studies at the
  LHC in the Standard Model}},  {\em Pos} {\bf DIS2014} (2014) 120,
  [\href{http://xxx.lanl.gov/abs/1407.1045}{{\tt arXiv:1407.1045}}].

\bibitem{Plehn:2001nj}
T.~Plehn, D.~L. Rainwater, and D.~Zeppenfeld, {\it {Determining the structure
  of Higgs couplings at the LHC}},  {\em Phys. Rev. Lett.} {\bf 88} (2002)
  051801, [\href{http://xxx.lanl.gov/abs/hep-ph/0105325}{{\tt
  hep-ph/0105325}}].

\bibitem{Hankele:2006ma}
V.~Hankele, G.~Klamke, D.~Zeppenfeld, and T.~Figy, {\it {Anomalous Higgs boson
  couplings in vector boson fusion at the CERN LHC}},  {\em Phys. Rev.} {\bf
  D74} (2006) 095001, [\href{http://xxx.lanl.gov/abs/hep-ph/0609075}{{\tt
  hep-ph/0609075}}].

\bibitem{Klamke:2007cu}
G.~Klamke and D.~Zeppenfeld, {\it {Higgs plus two jet production via gluon
  fusion as a signal at the CERN LHC}},  {\em JHEP} {\bf 04} (2007) 052,
  [\href{http://xxx.lanl.gov/abs/hep-ph/0703202}{{\tt hep-ph/0703202}}].

\bibitem{Hagiwara:2009wt}
K.~Hagiwara, Q.~Li, and K.~Mawatari, {\it {Jet angular correlation in
  vector-boson fusion processes at hadron colliders}},  {\em JHEP} {\bf 07}
  (2009) 101, [\href{http://xxx.lanl.gov/abs/0905.4314}{{\tt
  arXiv:0905.4314}}].

\bibitem{Buckley:2010jv}
M.~R. Buckley and M.~J. Ramsey-Musolf, {\it {Diagnosing Spin at the LHC via
  Vector Boson Fusion}},  {\em JHEP} {\bf 09} (2011) 094,
  [\href{http://xxx.lanl.gov/abs/1008.5151}{{\tt arXiv:1008.5151}}].

\bibitem{Campanario:2010mi}
F.~Campanario, M.~Kubocz, and D.~Zeppenfeld, {\it {Gluon-fusion contributions
  to Phi + 2 Jet production}},  {\em Phys. Rev.} {\bf D84} (2011) 095025,
  [\href{http://xxx.lanl.gov/abs/1011.3819}{{\tt arXiv:1011.3819}}].

\bibitem{Hagiwara:2013jp}
K.~Hagiwara and S.~Mukhopadhyay, {\it {Azimuthal correlation among jets
  produced in association with a bottom or top quark pair at the LHC}},  {\em
  JHEP} {\bf 05} (2013) 019, [\href{http://xxx.lanl.gov/abs/1302.0960}{{\tt
  arXiv:1302.0960}}].

\bibitem{Budnev:1974de}
V.~M. Budnev, I.~F. Ginzburg, G.~V. Meledin, and V.~G. Serbo, {\it {The Two
  photon particle production mechanism. Physical problems. Applications.
  Equivalent photon approximation}},  {\em Phys. Rept.} {\bf 15} (1975)
  181--281.

\bibitem{Bambade:1993vw}
P.~Bambade, A.~Dobrovolskaya, and V.~Novikov, {\it {Azimuthal asymmetry in
  Higgs production via vector boson fusion}},  {\em Phys. Lett.} {\bf B319}
  (1993) 348--354.

\bibitem{DelDuca:2003ba}
V.~Del~Duca, W.~Kilgore, C.~Oleari, C.~R. Schmidt, and D.~Zeppenfeld, {\it
  {Kinematical limits on Higgs boson production via gluon fusion in association
  with jets}},  {\em Phys. Rev.} {\bf D67} (2003) 073003,
  [\href{http://xxx.lanl.gov/abs/hep-ph/0301013}{{\tt hep-ph/0301013}}].

\bibitem{Figy:2004pt}
T.~Figy and D.~Zeppenfeld, {\it {QCD corrections to jet correlations in weak
  boson fusion}},  {\em Phys. Lett.} {\bf B591} (2004) 297--303,
  [\href{http://xxx.lanl.gov/abs/hep-ph/0403297}{{\tt hep-ph/0403297}}].

\bibitem{Campbell:2006xx}
J.~M. Campbell, R.~K. Ellis, and G.~Zanderighi, {\it {Next-to-Leading order
  Higgs + 2 jet production via gluon fusion}},  {\em JHEP} {\bf 10} (2006) 028,
  [\href{http://xxx.lanl.gov/abs/hep-ph/0608194}{{\tt hep-ph/0608194}}].

\bibitem{DelDuca:2006hk}
V.~Del~Duca, G.~Klamke, D.~Zeppenfeld, M.~L. Mangano, M.~Moretti, F.~Piccinini,
  R.~Pittau, and A.~D. Polosa, {\it {Monte Carlo studies of the jet activity in
  Higgs + 2 jet events}},  {\em JHEP} {\bf 10} (2006) 016,
  [\href{http://xxx.lanl.gov/abs/hep-ph/0608158}{{\tt hep-ph/0608158}}].

\bibitem{Andersen:2007mp}
J.~R. Andersen, T.~Binoth, G.~Heinrich, and J.~M. Smillie, {\it {Loop induced
  interference effects in Higgs Boson plus two jet production at the LHC}},
  {\em JHEP} {\bf 02} (2008) 057,
  [\href{http://xxx.lanl.gov/abs/0709.3513}{{\tt arXiv:0709.3513}}].

\bibitem{Ciccolini:2007ec}
M.~Ciccolini, A.~Denner, and S.~Dittmaier, {\it {Electroweak and QCD
  corrections to Higgs production via vector-boson fusion at the LHC}},  {\em
  Phys. Rev.} {\bf D77} (2008) 013002,
  [\href{http://xxx.lanl.gov/abs/0710.4749}{{\tt arXiv:0710.4749}}].

\bibitem{Andersen:2008gc}
J.~R. Andersen, V.~Del~Duca, and C.~D. White, {\it {Higgs Boson Production in
  Association with Multiple Hard Jets}},  {\em JHEP} {\bf 02} (2009) 015,
  [\href{http://xxx.lanl.gov/abs/0808.3696}{{\tt arXiv:0808.3696}}].

\bibitem{Bredenstein:2008tm}
A.~Bredenstein, K.~Hagiwara, and B.~{J\"{a}ger}, {\it {Mixed QCD-electroweak
  contributions to Higgs-plus-dijet production at the LHC}},  {\em Phys. Rev.}
  {\bf D77} (2008) 073004, [\href{http://xxx.lanl.gov/abs/0801.4231}{{\tt
  arXiv:0801.4231}}].

\bibitem{Nason:2009ai}
P.~Nason and C.~Oleari, {\it {NLO Higgs boson production via vector-boson
  fusion matched with shower in POWHEG}},  {\em JHEP} {\bf 02} (2010) 037,
  [\href{http://xxx.lanl.gov/abs/0911.5299}{{\tt arXiv:0911.5299}}].

\bibitem{Andersen:2010zx}
J.~R. Andersen, K.~Arnold, and D.~Zeppenfeld, {\it {Azimuthal Angle
  Correlations for Higgs Boson plus Multi-Jet Events}},  {\em JHEP} {\bf 06}
  (2010) 091, [\href{http://xxx.lanl.gov/abs/1001.3822}{{\tt
  arXiv:1001.3822}}].

\bibitem{Campanario:2013mga}
F.~Campanario and M.~Kubocz, {\it {Higgs boson production in association with
  three jets via gluon fusion at the LHC: Gluonic contributions}},  {\em Phys.
  Rev.} {\bf D88} (2013), no.~5 054021,
  [\href{http://xxx.lanl.gov/abs/1306.1830}{{\tt arXiv:1306.1830}}].

\bibitem{Campanario:2014oua}
F.~Campanario and M.~Kubocz, {\it {Higgs boson CP-properties of the gluonic
  contributions in Higgs plus three jet production via gluon fusion at the
  LHC}},  {\em JHEP} {\bf 10} (2014) 173,
  [\href{http://xxx.lanl.gov/abs/1402.1154}{{\tt arXiv:1402.1154}}].

\bibitem{DelDuca:2001eu}
V.~Del~Duca, W.~Kilgore, C.~Oleari, C.~Schmidt, and D.~Zeppenfeld, {\it {Higgs
  + 2 jets via gluon fusion}},  {\em Phys. Rev. Lett.} {\bf 87} (2001) 122001,
  [\href{http://xxx.lanl.gov/abs/hep-ph/0105129}{{\tt hep-ph/0105129}}].

\bibitem{DelDuca:2001fn}
V.~Del~Duca, W.~Kilgore, C.~Oleari, C.~Schmidt, and D.~Zeppenfeld, {\it {Gluon
  fusion contributions to H + 2 jet production}},  {\em Nucl. Phys.} {\bf B616}
  (2001) 367--399, [\href{http://xxx.lanl.gov/abs/hep-ph/0108030}{{\tt
  hep-ph/0108030}}].

\bibitem{Alwall:2014hca}
J.~Alwall, R.~Frederix, S.~Frixione, V.~Hirschi, F.~Maltoni, O.~Mattelaer,
  H.~S. Shao, T.~Stelzer, P.~Torrielli, and M.~Zaro, {\it {The automated
  computation of tree-level and next-to-leading order differential cross
  sections, and their matching to parton shower simulations}},  {\em JHEP} {\bf
  07} (2014) 079, [\href{http://xxx.lanl.gov/abs/1405.0301}{{\tt
  arXiv:1405.0301}}].

\bibitem{Hirschi:2015iia}
V.~Hirschi and O.~Mattelaer, {\it {Automated event generation for loop-induced
  processes}},  {\em JHEP} {\bf 10} (2015) 146,
  [\href{http://xxx.lanl.gov/abs/1507.0002}{{\tt arXiv:1507.0002}}].

\bibitem{Binosi:2003yf}
D.~Binosi and L.~Theussl, {\it {JaxoDraw: A Graphical user interface for
  drawing Feynman diagrams}},  {\em Comput. Phys. Commun.} {\bf 161} (2004)
  76--86, [\href{http://xxx.lanl.gov/abs/hep-ph/0309015}{{\tt
  hep-ph/0309015}}].

\bibitem{Williams:1934ad}
E.~J. Williams, {\it {Nature of the high-energy particles of penetrating
  radiation and status of ionization and radiation formulae}},  {\em Phys.
  Rev.} {\bf 45} (1934) 729--730.

\bibitem{vonWeizsacker:1934nji}
C.~F. von Weizsacker, {\it {Radiation emitted in collisions of very fast
  electrons}},  {\em Z. Phys.} {\bf 88} (1934) 612--625.

\bibitem{Hagiwara:1985yu}
K.~Hagiwara and D.~Zeppenfeld, {\it {Helicity Amplitudes for Heavy Lepton
  Production in e+ e- Annihilation}},  {\em Nucl. Phys.} {\bf B274} (1986)
  1--32.

\bibitem{Hagiwara:1988pp}
K.~Hagiwara and D.~Zeppenfeld, {\it {Amplitudes for Multiparton Processes
  Involving a Current at e+ e-, e+- p, and Hadron Colliders}},  {\em Nucl.
  Phys.} {\bf B313} (1989) 560--594.

\bibitem{Pumplin:2002vw}
J.~Pumplin, D.~R. Stump, J.~Huston, H.~L. Lai, P.~M. Nadolsky, and W.~K. Tung,
  {\it {New generation of parton distributions with uncertainties from global
  QCD analysis}},  {\em JHEP} {\bf 07} (2002) 012,
  [\href{http://xxx.lanl.gov/abs/hep-ph/0201195}{{\tt hep-ph/0201195}}].

\bibitem{Kawabata:1995th}
S.~Kawabata, {\it {A New version of the multidimensional integration and event
  generation package BASES/SPRING}},  {\em Comput. Phys. Commun.} {\bf 88}
  (1995) 309--326.

\bibitem{vanOldenborgh:1989wn}
G.~J. van Oldenborgh and J.~A.~M. Vermaseren, {\it {New Algorithms for One Loop
  Integrals}},  {\em Z. Phys.} {\bf C46} (1990) 425--438.

\bibitem{Shifman:1979eb}
M.~A. Shifman, A.~I. Vainshtein, M.~B. Voloshin, and V.~I. Zakharov, {\it
  {Low-Energy Theorems for Higgs Boson Couplings to Photons}},  {\em Sov. J.
  Nucl. Phys.} {\bf 30} (1979) 711--716. [Yad. Fiz.30,1368(1979)].

\bibitem{Degrande:2011ua}
C.~Degrande, C.~Duhr, B.~Fuks, D.~Grellscheid, O.~Mattelaer, and T.~Reiter,
  {\it {UFO - The Universal FeynRules Output}},  {\em Comput. Phys. Commun.}
  {\bf 183} (2012) 1201--1214, [\href{http://xxx.lanl.gov/abs/1108.2040}{{\tt
  arXiv:1108.2040}}].

\bibitem{Christensen:2008py}
N.~D. Christensen and C.~Duhr, {\it {FeynRules - Feynman rules made easy}},
  {\em Comput. Phys. Commun.} {\bf 180} (2009) 1614--1641,
  [\href{http://xxx.lanl.gov/abs/0806.4194}{{\tt arXiv:0806.4194}}].

\bibitem{Catani:2001cc}
S.~Catani, F.~Krauss, R.~{K\"uhn}, and B.~R. Webber, {\it {QCD matrix elements
  + parton showers}},  {\em JHEP} {\bf 11} (2001) 063,
  [\href{http://xxx.lanl.gov/abs/hep-ph/0109231}{{\tt hep-ph/0109231}}].

\bibitem{Lonnblad:2001iq}
L.~Lonnblad, {\it {Correcting the color dipole cascade model with fixed order
  matrix elements}},  {\em JHEP} {\bf 05} (2002) 046,
  [\href{http://xxx.lanl.gov/abs/hep-ph/0112284}{{\tt hep-ph/0112284}}].

\bibitem{Alwall:2007fs}
J.~Alwall {\em et.~al.}, {\it {Comparative study of various algorithms for the
  merging of parton showers and matrix elements in hadronic collisions}},  {\em
  Eur. Phys. J.} {\bf C53} (2008) 473--500,
  [\href{http://xxx.lanl.gov/abs/0706.2569}{{\tt arXiv:0706.2569}}].

\bibitem{Nakamura:2015uca}
J.~Nakamura, {\it {A simple merging algorithm for jet angular correlation
  studies}},  \href{http://xxx.lanl.gov/abs/1509.0416}{{\tt arXiv:1509.0416}}.

\end{thebibliography}\endgroup

\end{document}